\documentclass[10pt]{article}
\usepackage[OT1]{fontenc}
\pdfoutput=1
\usepackage[dvipsnames]{xcolor}
\usepackage{amsthm,amsmath,amsbsy,amssymb,array,bm,graphicx,epstopdf,mathtools,calc,xfrac}
\usepackage[sort]{natbib}
\usepackage{xr-hyper}
\RequirePackage[activate={true,nocompatibility},final,tracking=true,kerning=true,spacing=true,factor=1100,stretch=10,shrink=10,spacing=basictext]{microtype}
\microtypecontext{spacing=nonfrench}
\usepackage{authblk}

\usepackage{url} 
\usepackage[ruled,linesnumbered]{algorithm2e}
\SetKwInput{KwInput}{Input}
\SetKwInput{KwOutput}{Output}
\usepackage{subfigure,subdepth,mdframed,tikz,pgf,pgffor}
\usepgfmodule{shapes,plot}
\usetikzlibrary{decorations,arrows,positioning,calc}
\xdefinecolor{darkgreen}{RGB}{0,193,0}
 
 \newlength{\mylength}
\allowdisplaybreaks
\makeatletter
\newcommand{\pushright}[1]{\ifmeasuring@#1\else\omit\hfill$\displaystyle#1$\fi\ignorespaces}
\newcommand{\pushleft}[1]{\ifmeasuring@#1\else\omit$\displaystyle#1$\hfill\fi\ignorespaces}
\makeatother
\let\originalleft\left%
\let\originalright\right%
\renewcommand{\left}{\mathopen{}\mathclose\bgroup\originalleft}%
\renewcommand{\right}{\aftergroup\egroup\originalright}%
\makeatletter%
\def\mybigx#1{\dimen@#1\relax%
\mathchoice%
{\vbox to \dimen@{}}%
{\vbox to \dimen@{}}%
{\vbox to .7\dimen@{}}%
{\vbox to .5\dimen@{}}}%
\def\mybig#1{{\hbox{$\left#1\mybigx{0.8em}\right.\n@space$}}}%
\makeatother%
\usepackage[titletoc,title]{appendix}
\usepackage{subfigure}
\usepackage{lipsum} 
\usepackage{a4wide}
\usepackage{float}
\usepackage{xcolor}
\usepackage{amsthm}
\usepackage{graphicx}
\usepackage[latin1]{inputenc}
\usepackage{multirow} 
\usepackage{bm}
\usepackage{epsfig}
\usepackage{lscape}
\usepackage{bm}

\DeclareFontFamily{U}{mathx}{\hyphenchar\font45}
\DeclareFontShape{U}{mathx}{m}{n}{<-> mathx10}{}
\DeclareSymbolFont{mathx}{U}{mathx}{m}{n}
\usepackage{thmtools}
\usepackage{thm-restate}

\usepackage[noabbrev,capitalise,nameinlink]{cleveref}
\numberwithin{equation}{section}
\theoremstyle{plain}
\crefname{figure}{Fig.}{Figs.}
\crefname{table}{Table}{Tables}
\newtheorem{Definition}{Definition}
\crefname{Definition}{Definition}{Definition}
\crefname{defin}{Definition}{Definition}

\crefname{Lemma}{Lemma}{Lemmas}

\crefname{Proposition}{Proposition}{Propositions}

\crefname{Theorem}{Theorem}{Theorems}
\crefname{thm}{Theorem}{Theorems}

\crefname{Corollary}{Corollary}{Corollaries}
\theoremstyle{remark}

\crefname{Remark}{Remark}{Remarks}
\newcommand{\bbeta}{\boldsymbol{\beta}}

\newcommand{\bmu}{\boldsymbol{\mu}}

\newcommand{\bgamma}{\boldsymbol{\gamma}}
\newcommand{\balpha}{\boldsymbol{\alpha}}

\newcommand{\vecc}{\text{vec}}

\newcommand{\argmin}{\arg\!\min}

\usepackage{etoolbox}

\usepackage{color}
\definecolor{darklavender}{rgb}{0.45, 0.31, 0.59}

\newcommand{\iid}{ \stackrel{i.i.d.}{=} }

\makeatletter 
   \newtheoremstyle{Example}{\topsep}{\topsep}%
     {}
     {}
     {\bfseries}
     {:}
    {0.9mm}
     {\thmname{#1}\thmnumber{ #2}\thmnote{(\it #3)}}
		
   \theoremstyle{Example}
   
	\AtEndEnvironment{Example}{\null\hfill\qed}%
\makeatother

\begin{document}
\title{\bf Profile least squares estimation of structure in networks with covariates}
\author[1*]{Swati Chandna}
\author[2]{Benjamin Bagozzi}
\author[3]{Snigdhansu Chatterjee}
\affil[1]{School of Computing and Mathematical Sciences, Birkbeck, University of London, U.K.}
\affil[2]{University of Delaware, USA}
\affil[3]{University of Maryland, Baltimore County, USA}
\affil[*]{s.chandna@bbk.ac.uk}
\date{}
 \maketitle
\begin{abstract}
	Many real world networks exhibit edge heterogeneity with different pairs of nodes interacting with different intensities. Further, nodes with similar attributes tend to interact more with each other. 
		Thus, in the presence of observed node attributes (covariates), it is of interest to understand the extent to which these covariates explain interactions between pairs of nodes and to suitably estimate the remaining structure due to unobserved factors. For example, in the study of international relations,
		the extent to which country-pair specific attributes such as the number of material/verbal conflicts and volume of trade
		explain military alliances between different countries can lead to valuable insights. 
		We study the model where pairwise edge probabilities are given by the sum of a linear edge covariate term and a residual term to model the remaining heterogeneity from unobserved factors.
		We approach estimation of the model via profile least squares and show
		how it leads to a simple algorithm to estimate the linear covariate term and the residual structure that is truly latent in the presence of observed covariates.
		Our framework lends itself naturally to a bootstrap procedure which is used to draw inference on model parameters, such as to determine significance of the homophily parameter or covariates in explaining the underlying network structure. Application to four real network datasets and comparisons using simulated data illustrate the usefulness of our approach.
\end{abstract}
\section{Introduction}
Real networks are characterized by node heterogeneity and edge homophily where nodes sharing common attributes tend to interact more with each other. 
Most economic and social science applications routinely record network data with attributes at the node or edge level, which at least partly explains the popularity of statistical network models that allow integration of covariates such as \cite{hoff02}, \cite{Choi2012}, \cite{Sweet15}, \cite{Latouche15}, \cite{Rohe17}, \cite{Huang2018} and more recently \cite{Muetal}, \cite{Chandna2021local} and \cite{Su2020}, to name a few. Specifically, there has been a growing interest in modeling edge homophily due to covariates, 
and in explaining edge formation via latent position models, such as \cite{Mele2023}, \cite{Roy2019}, \cite{Graham17}, \cite{dzemski2019empirical}, \cite{yan2019}, \cite{hoff02}.
The increasing complexity of networks in practice necessitates the need for models which can capture the key features and complexity as observed in real network data, and can be easily estimated using classical statistical methods. 
With the objective of achieving this goal, we propose the model where pairwise edge probabilities are determined by a linear covariate term and a generalized inner product of latent positions in a low-dimensional Euclidean space to model heterogeneity due to unobserved factors.
Our motivation to consider this model is inspired by the analogy we make  with the well-known partially linear model which has been studied extensively in classical semiparametric statistics (e.g. \cite{Hardle2000partially}, \cite{Fan2005profile}), and the relatively more recent developments in spectral estimation of structure in networks under the random dot product graph (RDPG), \cite{Athreya2017},
and the generalized random dot product graph (GRDPG), \cite{Rubin2022}. 
We study profile least squares estimation of the proposed model and establish why its extension to networks does not follow in the same way as in classical semiparametric statistics (e.g. \cite{Speckman1988}).
Subsequently, we show how the profile framework can be utilized to develop a simple iterative profile least squares algorithm for estimation of the model. 
Our algorithm is based on well-known existing methods and thus provides a simple approach to estimating and visualizing covariate contribution to the overall network topology (e.g. covariates explaining node-level heterogeneity and/or leading to clusters in the overall network structure) and to test 
their significance.

The data comprises of a single undirected binary network on $n$ nodes, represented via the symmetric adjacency matrix $A=(A_{ij})_{n\times n}$ with $A_{ij}=1$ (or $0$) denoting the presence (or absence) of interaction between nodes $i$ and $j$. We assume that there are no self-loops and hence $A_{ii}=0$.
With $\{\balpha_{i}\}_{i=1}^{n},$ $\balpha_{i}\in \mathbb{R}^{d}$ denoting the latent node-specific vectors, consider the model where interactions $A_{ij}, i<j$ arise as conditionally independent Bernoulli trials given $\balpha_{i},\balpha_{j}$
with edge probability $P_{ij}=Pr(A_{ij}=1)$, modelled as a linear function of edge covariates and a nonparametric function of latent vectors, i.e.
\begin{equation}\label{eqn:logitun}
	g(P_{ij})= x^{T}_{ij}\gamma + f(\balpha_{i},\balpha_{j}),
\end{equation}
where $g$ denotes the link function (e.g. logit), $x_{ij}=[x_{1,ij},\ldots,x_{p,ij}]^{T}$ is a $p \times 1$ vector of edge covariates observed for node pair $(i,j)$; $\gamma=[\gamma_{1},\ldots,\gamma_{p}]^{T}$ is the corresponding homophily parameter; and $f$ is some function of latent node-specific vectors $\balpha_{i}$ and $\balpha_{j}$. 
Assuming $f$ to be of a specific form makes the problem easier and a variety of forms have been proposed in the literature (summarized in \cref{sec:sublitreview} below). In this paper, we study another novel variant of the model in \eqref{eqn:logitun} where we employ the indefinite inner product kernel following the generalized random dot product graph model (GRDPG) by \cite{Rubin2022}, to model the pairwise residual  structure in the presence of the linear covariate term and with $g$ as the identity link. This model has the same form as the GRDPG with vertex covariates model introduced in \cite{Muetal} 
however, with two key differences. First, the linear term in their model is restricted to discrete categorical covariates.
In contrast, our model allows inclusion of continuous and/or categorical covariates. Additionally, the focus of their work is on community detection in network data with vertex covariates and hence they specifically study the special case where the indefinite inner product kernel corresponds to a stochastic blockmodel (SBM). We do not work under this specific assumption and are mainly concerned with estimation and inference on covariate effects through the linear term and 
their contribution to the overall heterogeneity in network structure. 

We approach estimation of the proposed model using the profile least squares (PLS) method which has been a natural choice for estimation of classical semiparametric models such as the partially linear model, \cite{Speckman1988}
where likewise, interest lies in the parametric component given by the linear covariate term. The use of PLS for our setting is motivated by the observation that with $g$ as the identity link, \eqref{eqn:logitun} may be viewed as a network analogue of the well-known partially linear model. This analogy also immediately highlights the main difference between our setting of unlabeled networks and the classical semiparametric setting. Specifically, the latency of node specific design points $\balpha_{i}, i \in \{1,\ldots,n\}$, in networks makes the application of classical approaches (such as of \cite{Speckman1988}) to corresponding network models (such as \eqref{eqn:logitun}) particularly challenging. If the node-specific vectors $\balpha_{i}$ were observed, one could easily estimate the model using classical profile least squares, e.g. \cite{Speckman1988}. Since this is not the case with unlabeled network data, simplifying assumptions such as assuming specific functional forms for the nonparametric component $f$ are crucial to estimation in this setting. Noting this and with the aim of providing a computationally simple, frequentist approach rooted in classical profile least squares estimation, we work with $f$ in \eqref{eqn:logitun} as the indefinite inner product kernel of the GRDPG model, \cite{Rubin2022}.
Note that although the form of the function $f$ is specified via the GRDPG kernel, the `nonparametric' term $f(\balpha_{i},\balpha_{j})$ in our model remains unknown due to the latency of node-specific vector positions $\balpha_{i}, i \in [n]$, including their dimension $d$. 
We study profile least squares estimation of this model and show why it does not lead to a closed-form estimator of the linear covariate coefficient as in the classical setting.
Subsequently, following this framework, an iterative profile least squares algorithm is proposed. 
Further, we show how the generalized bootstrap of \cite{Chatterjee2005generalized} lends itself naturally to this framework and makes inference on the unknown parameters of the model feasible.
This provides a powerful tool for inference on the model parameters, especially when small to medium sized networks are observed.

In practice, our method allows estimation and visualization of the covariate effect and the residual term which is useful to understand how covariates contribute to the overall network structure. The usefulness of our method 
is illustrated by application to a variety of real networks observing continuous only or continuous and categorical covariates.
For example, in one data example (tree network), we observe significant positive values in the residual kernel, with covariates only explaining heterogeneity in the network at the node-level;
whereas in another example (CKM physician friendship), covariates clearly contribute to formation of clusters in the network, with the residual kernel being close to zero for a large subset of node pairs. We observe less of a contrast in the other two data examples (military alliance networks), where our edge probability estimator has clusters purely due to observed covariates for a subset of node pairs and is either a sum of contributions from both the observed covariate and the residual term (unobserved factors) or only from the residual term, for the remaining node pairs. 
To summarize, our method allows a fine grained analysis of the structure in the network by providing a clear decomposition of the network topology (modelled via the edge probability matrix) in terms of the contribution from the observed characteristics (through the linear covariate term) and the unobserved latent characteristics (through the residual kernel).

\subsection{Related literature}\label{sec:sublitreview}
As discussed above, assuming $f$ in \eqref{eqn:logitun} to be of a specific form makes the problem easier and a variety of forms have been proposed and studied in the literature, some of which we summarize below.
The well-known $\beta$-model of \cite{Chatterjee11AoP} studied \eqref{eqn:logitun} with $g$ as the logit link, without the covariate term ($\gamma=0$) and with $f(\alpha_{i},\alpha_{j})=\alpha_{i}+\alpha_{j}$, thus assuming scalar latent positions. This has been the basis of subsequent contributions allowing the inclusion of covariates via a linear term as in \eqref{eqn:logitun}. For example, as in \cite{Graham17} and \cite{dzemski2019empirical}, where the latent additive term $\alpha_{i}+\alpha_{j}$ models degree heterogeneity, and focus is on estimation of the homophily parameter $\gamma$. The $\beta$-model with covariates for directed networks as in \cite{yan2019} modifies the functional form to $f(\alpha_{i},\beta_{j})= \alpha_{i} + \beta_{j}$, with $\alpha_{i}$ determining the outgoingness of node $i$, and $\beta_{j}$ denoting the incoming behavior or popularity of node $j$.
In the well-known class of latent space models (\cite{hoff02}, \cite{Krivitsky2009}), the pairwise interaction between nodes is modelled again as a special case of \eqref{eqn:logitun} with $f$ specified to correspond to a distance, e.g.~with $f(\balpha_{i},\balpha_{j})=||\balpha_{i}-\balpha_{j}||_{2}$ (Euclidean distance) and $f(\balpha_{i},\balpha_{j})=\balpha^{T}_{i}\balpha_{j}$ (bilinear distance) in \cite{hoff02}.
Another category of contributions comprise those
where $f$ specifically takes the form of a SBM. This is not surprising given the popularity of the community detection problem. The basic assumption underlying these models is that covariates influence the probability of an edge independently of the block membership. 
This setting was originally studied by \cite{Choi2012} and subsequently formally introduced by \cite{Sweet15} as Covariate Stochastic BlockModels (CSBMs) who proposed a bayesian model fitting procedure. More recently,
\cite{Roy2019} proposed an EM type algorithm based on case-control approximation of the log-likelihood. 
Two latest contributions in this category of methods are \cite{Mele2023} and \cite{Muetal}, where $f$ specifically takes the form of the SBM as a special case of the random dot product kernel, and the generalized random dot product kernel, respectively. \cite{Mele2023} employs $g$ as the logit link whereas \cite{Muetal} studied a linear probability model. 

A related problem concerns determining the significance of covariates in explaining the whole topology of the network.
With a view to assess the goodness of fit of logistic regression models, the Bayesian approach of \cite{Latouche15} studies \eqref{eqn:logitun} with $f$ as the general nonparametric graphon function. Given the intractibility of the graphon function, their original model is approximated with a series of models with blockwise constant structures. Thus, an instance of their approximation corresponds to
contributions employing SBMs to capture the latent network structure (such as \cite{Choi2012}, \cite{Sweet15}, \cite{Roy2019}, \cite{Mele2023}, \cite{Muetal}), as also noted in \cite{Roy2019}.
\section{Modeling edge probabilities}\label{sec:model}
We consider another variant of \eqref{eqn:logitun} with $g$ as the identity link and with the residual structure modelled through $f$ specified as the indefinite inner product kernel, \cite{Rubin2022}, which is defined below.
\begin{Definition}[Indefinite inner product kernel, \cite{Rubin2022}]
	For vectors $\mathbf{x}\in \mathbb{R}^{q+s}$ and $\mathbf{y}\in \mathbb{R}^{q+s}$, the indefinite inner product kernel, denoted as $f(\mathbf{x},\mathbf{y};q,s)$ is given by 
	\begin{align}\label{eqn:grdpgker}
		f(\mathbf{x},\mathbf{y};q,s)=\mathbf{x}^{T}\mathbb{I}_{qs}\mathbf{y}&=x_{1}y_{1}+\ldots+x_{q}y_{q}\nonumber\\
		&-x_{q+1}y_{q+1}-\ldots -x_{q+s}y_{q+s},
	\end{align}
	where $\mathbb{I}_{qs}$ is a block diagonal matrix given by 
	\begin{equation*}
		\mathbb{I}_{qs}=\begin{bmatrix}\mathbb{I}_{q} & \mathbf{0}\\
			\mathbf{0} & -\mathbb{I}_{s}\end{bmatrix},
	\end{equation*} 
	with $\mathbb{I}_{r}$ for any positive $r$ representing the $r \times r$ identity matrix.
\end{Definition}\label{def:grdpg}

The GRDPG model introduced in \cite{Rubin2022} is a generalization of the well-known random dot-product graph (RDPG) model, \cite{Young2007}, and specifies an indefinite inner product kernel to represent the probabilities of pairwise interactions in a network. It is key to modeling graph connectivity structures such as heterophilic connectivity (opposites attract) and core-periphery connectivity, that the RDPG model does not allow.
We use the indefinite inner product kernel \eqref{eqn:grdpgker} to allow for flexibility in modeling the residual  structure.
This is crucial in the presence of covariates since the residual structure may or may not be limited to edge homophily  depending on the observed set of covariates.
For example, in a student friendship network, students within the same grade are typically more likely to interact and thus if student grades are observed, they may be employed to explain edge homophily. However, if student grades are the only observed covariate, strong interactions between students in different grades (e.g. from participation in other school activities, ethnicity or any other unobserved attributes) is edge heterophily (w.r.t grades) and hence must be captured by the model for the residual structure. With this view, we have chosen the indefinite inner product kernel to allow representation of varied residual structures in the presence of covariates, such as residual heterophily. Our network-covariate model is formally defined below.
\begin{Definition}
	Given $d\geq 1$ dimensional latent position vectors $\balpha_{i}=[\alpha_{i1},\ldots,\alpha_{id}]^{T}, i \in \{1,\ldots,n\}\equiv [n]$ and edge covariates $x_{ij}=[x_{1,ij},\ldots,x_{p,ij}]^{T}$, we model $A_{ij} \in\{0,1\}$ as conditionally independent Bernoulli trials with success probability $P_{ij}$, $(i,j)\in [n]\times [n]$ where, with $\bgamma=[\gamma_{1},\ldots,\gamma_{p}]^{T}$, for $i<j$, we have
	\begin{equation}\label{eqn:RDPGcov}
		P_{ij}= x^{T}_{ij}\bgamma + f(\balpha_{i},\balpha_{j};q,s),
	\end{equation}
	where, $f$ denotes the indefinite inner product kernel as in \eqref{eqn:grdpgker}, for integer-valued $q>0$ and $s\geq 0$ such that $q+s=d$.
\end{Definition}
Thus, model \eqref{eqn:RDPGcov} associates to each node $i$, a $d$-dimensional vector parameter $\balpha_{i}$, such that the residual structure is modelled through the pairwise indefinite inner product. Here the first $q$ dimensions are referred to as \textit{assortative} whereas the remaining $s$ dimensions are \textit{disassortative}, \cite{Rubin2022}. In the context of our model, $s\geq 1$ corresponds to the case where covariates lead to residual heterophily in contrast to the case of $s=0$ which implies that the indefinite inner product kernel $f$ is only used to capture residual homophily and hence equivalent to employing the pairwise dot product kernel $\balpha^{T}_{i}\balpha_{j}$ (e.g. as used under the RDPG model) for the residual term. 

We consider two types of residual structures that may arise in practice under our framework. The first is the setting where the residual term represented by $f$ in \eqref{eqn:RDPGcov} corresponds to the well-known stochastic blockmodel, 
as, for example, studied in \cite{Choi2012}, \cite{Sweet15}, \cite{Roy2019}, \cite{Muetal}, and \cite{Mele2023}. The second setting we consider is the case where the kernel $f$ in \eqref{eqn:RDPGcov} represents a general low-rank $K$-block structure which does not correspond to a valid network model.
We note that, in general, the RDPG and GRDPG models employing the dot product and indefinite inner product kernels, respectively, need not lead to valid edge probabilities in the range $[0,1]$. In fact, this was the main motivation for the assumption referred to as the \textit{inner product condition} by \cite{Young2007} and employed in subsequent literature on the topic (e.g. \cite{Athreya2017}). Since our model employs the indefinite inner product kernel $f$ to represent the residual structure and not edge probabilities of the network, it is natural to not impose this condition and allow the kernel to assume values outside the range $[0,1]$,
where required for a better overall fit.
We describe these two types of residual structures below.
\subsection*{Type I. ($K$-block SBM) }
Consider the setting where the residual term $f$ corresponds to
the well-known stochastic blockmodel with $K$ blocks. 
Given a block assignment vector $z=[z_{1},\ldots,z_{n}]^{T}$ and $K$ latent position vectors $\bmu_{1},\ldots,\bmu_{K} \in \mathbb{R}^{d}$ (where $d\leq K$), we have $\balpha_{i}=\bmu_{a}$ if $z_{i}=a$, for each $i \in [n]$ and $a\in[K]$. Then $f(\balpha_{i},\balpha_{j};q,s)=\Theta_{z_{i}z_{j}}$, where $\Theta_{ab}=\mu^{T}_{a}\mathbb{I}_{qs}\mu_{b}$ for $(a,b) \in [K]\times [K].$
Then, for example, with a single ($p=1$) binary edge covariate $x_{ij}\in \{0,1\},$  with homophily parameter or linear coefficient $\gamma$ such that $\gamma+\Theta_{z_{i}z_{j}}\in [0,1]$  we get:  
\begin{equation*}
	P_{ij}=\begin{cases}
		\Theta_{z_{i}z_{j}} & \text{if } x_{ij}=0\\
		\gamma+\Theta_{z_{i}z_{j}} & \text{if } x_{ij}=1
	\end{cases}.
\end{equation*}
This example corresponds to the (only) case studied in \cite{Muetal}. In this work, we also consider estimation for the setting where continuous (or both continuous and categorical) covariates are observed. Note that in contrast to the example of a discrete binary-valued covariate above, a single continuous edge covariate $x_{ij}\in \mathbb{R},$ would correspond to edge probabilites of the form $P_{ij}=\gamma x_{ij}+\Theta_{z_{i}z_{j}}$. Clearly, there is additional edge heterogeneity resulting from the covariate term in this case since, in general, $x_{ij}$ may differ for every node pair $(i,j)\in [n]\times[n].$

\subsection*{Type II. (Low-rank structure with $K$ clusters) }
This generalizes the first setting above by allowing 
block constant residual matrices that do not correspond to a valid network model.
For example, with the $d$-dimensional latent positions $\balpha_{i}$ arising from a finite mixture of $K$ point masses, however, with the corresponding indefinite inner product kernel values not in $[0,1]$.
Specifically, here we have $\Theta=(\Theta_{z_{i}z_{j}}) \in \mathbb{R}^{K\times K}$ where $\Theta_{z_{i}z_{j}}$ follows from the form of the kernel as given above, with latent positions corresponding to the location of the point masses 
of the clusters to which node pair $(i,j)$ is mapped via $z$. The example below illustrates how this setting may arise naturally under the model \eqref{eqn:RDPGcov} including covariates.

\textit{Example}: To illustrate this setting, we consider the simple case of a rank-$2$ residual matrix with $K=2$ clusters. 
Suppose that  $q=1=s$, with the first dimension being assortative, so that diag$(\mathbb{I}_{qs})=(1,-1).$ Let $\bmu_{a}=[\mu_{a1},\mu_{a2}]^{T}$ denote the $d=2$ dimensional latent position for cluster $a\in\{1,2$\}. The residual matrix $\Theta$ then follows as $\Theta=[\bmu_{1},\bmu_{2}]^{T}\mathbb{I}_{qs} [\bmu_{1},\bmu_{2}]$
implying that $\Theta_{11}=\mu^{2}_{11}-\mu^{2}_{12}$; $\Theta_{22}=\mu^{2}_{21}-\mu^{2}_{22}$ and $\Theta_{12}=\mu_{11}\mu_{21}-\mu_{12}\mu_{22}=\Theta_{21}.$
Clearly, for a valid network model under \eqref{eqn:RDPGcov}, we must have
\begin{eqnarray}\label{eqn:const2}
	\max\{\Theta_{z_{i}z_{j}}+x^{T}_{ij}\bgamma\} &\leq& 1 \hspace{2mm}\text{ and } \nonumber\\ 
	\min\{\Theta_{z_{i}z_{j}}+x^{T}_{ij}\bgamma\} &\geq& 0.
\end{eqnarray}
In general, the overall covariate contribution given by the term $x^{T}_{ij}\bgamma$ may assume values outside $[0,1]$.
Then, from \eqref{eqn:const2}, it is clear that the residual matrix $\Theta$ may lie in $\mathbb{R}^{K\times K}$, in contrast to the first setting where it corresponded to an edge probability matrix with values in $[0,1]^{K\times K}$. For example, with $0.5\leq x^{T}_{ij}\bgamma\leq 1.2$ for $(i,j)$ mapped to residual cluster $(1,1)$, we must have $-0.5\leq \Theta_{11}\leq-0.2$ for it to lead to a valid network model i.e. satisfy the two conditions in \eqref{eqn:const2}.

Thus, in this second setting, we may interpret $\Theta$ as a low-rank residual matrix, with the magnitude of its elements representing the strength of residual pairwise structure in the presence of observed covariates. 
Simulations with concrete examples under this setting are included in \cref{sec:sim2}.

Note that, irrespective of the two settings mentioned above, the overall structure in the network implied by our model, is not necessarily of a blockmodel type, i.e. edge probabilities may vary depending on node pairs. 

\section{Profile least squares estimation}\label{sec:plsexact}
Clearly estimating the model involves estimating the $d$ dimensional latent node vectors together with the homophily vector $\bgamma \in \mathbb{R}^{p \times 1}$, i.e. the unknown parameters are \\
$\theta:=\{(\bgamma^{T}, \balpha^{T}_{1},\ldots,\balpha^{T}_{n})^{T}; (\bgamma^{T}, \balpha^{T}_{1},\ldots,\balpha^{T}_{n})^{T}\in \mathbb{R}^{p+nd}\}$.
Given $A_{ij}\sim \text{Bernoulli}(P_{ij})$ where $P_{ij}= x^{T}_{ij}\gamma + \balpha^{T}_{i}\mathbb{I}_{qs}\balpha_{j}$, we consider least squares estimation of the unknown parameter vector $\theta$ as defined above. Let $\Lambda=[\balpha^{T}_{1},\balpha^{T}_{2},\ldots,\balpha^{T}_{n}]^{T}\in \mathbb{R}^{n \times d}$ denote the matrix of d-dimensional latent vectors and $\mathbf{X}_{l}=[x_{l,ij}]$ denote the $n \times n$ pairwise edge covariate matrix, for each $l\in [p]$.
We define $\hat{\theta}=\argmin_{\bgamma,\Lambda}\mathcal{L}(\bgamma,\Lambda;\mathbf{X},A)$, where $\mathcal{L}$ denotes the least squares objective function, 
\begin{equation}\label{eqn:LSOrdpg}
	\mathcal{L}(\bgamma,\Lambda;\mathbf{X},A)=\sum_{i<j}\left(A_{ij}-x^{T}_{ij}\bgamma - \balpha^{T}_{i}\mathbb{I}_{qs}\balpha_{j} \right)^{2}.
\end{equation}
Since our primary interest lies in the covariate coefficient vector $\bgamma$, we shall proceed by profiling out the latent positions in $\Lambda$ by first estimating it in terms of the unknown coefficient $\bgamma$, denoted as $\hat{\Lambda}(\bgamma)$, and subsequently obtaining the least squares objective function in terms of $\bgamma$ alone.
Hence, following \eqref{eqn:LSOrdpg}, we define $Y_{ij}(\bgamma):=A_{ij}-x^{T}_{ij}\bgamma$, for a given $\bgamma$, so that the least squares function \eqref{eqn:LSOrdpg} becomes:
\begin{equation}\label{eqn:LSOrdpg1pls}
	\mathcal{L}(\Lambda;\bgamma,\mathbf{X},A)=\sum_{i<j}\left(Y_{ij}(\bgamma)- \balpha^{T}_{i}\mathbb{I}_{qs}\balpha_{j} \right)^{2}.
\end{equation}
Then from \cite{Rubin2022}, it follows that with the latent dimension $\hat{d}$ suitably determined, the matrix of latent vectors can be estimated as $\hat{\Lambda}=\hat{\mathbf{U}}|\hat{\mathbf{S}}|^{1/2} \in \mathbb{R}^{n \times \hat{d}}$ with $\hat{\mathbf{S}}$ as the diagonal matrix with the $\hat{d}$ largest (in magnitude) eigenvalues of $Y$,
on its diagonal, arranged in decreasing order (based on their actual, signed, value), and $\hat{\mathbf{U}}$ as the $n\times \hat{d}$ matrix with columns as the corresponding orthonormal eigenvectors arranged in the same order.  Thus, with $\hat{q}$ and $\hat{s}$ denoting the number of strictly positive and strictly negative eigenvalues of $Y$ such that $\hat{q}+\hat{s}=d$, we get
\begin{eqnarray}\label{eqn:lambhat}
	\hat{\Lambda}\mathbb{I}_{\hat{q}\hat{s}}\hat{\Lambda}^{T}&=&(\hat{\mathbf{U}}|\hat{\mathbf{S}}|^{1/2}) \mathbb{I}_{\hat{q}\hat{s}}(|\hat{\mathbf{S}}|^{1/2}\hat{\mathbf{U}}^{T}) \nonumber\\
	&=&\hat{\mathbf{U}}\hat{\mathbf{S}}\hat{\mathbf{U}}^{T}\\
	&=&\mathcal{F}Y(\bgamma)\nonumber,
\end{eqnarray}
where for any symmetric real-valued $n \times n$ matrix $\mathbf{B}$, we define $\mathcal{F}\mathbf{B}$ as the (low) rank-$\hat{d}$ filter operator as follows.
Let $\mathbf{B}=\mathbf{V}\mathbf{S}\mathbf{V}^{T}$ denote its eigendecomposition, where it is assumed that $\mathbf{S}$ is the diagonal matrix with the $\hat{d}$ largest (in magnitude) eigenvalues of $\mathbf{B}$,
on its diagonal, arranged in decreasing order (based on their actual, signed, value).
Then $\mathcal{F}\mathbf{B}=\mathbf{V}(\mathbf{F}\mathbf{S})\mathbf{V}^{T}$, where $\mathbf{F}=\begin{bmatrix} \mathbb{I}_{\hat{d}}  & \mathbf{0}\\
		\mathbf{0}& \mathbf{0}
	\end{bmatrix}.$
Thus, by definition, $\mathcal{F}$ applied to the covariate adjusted adjacency $Y(\bgamma)$ i.e. $\mathcal{F}Y$ leads to the low-rank approximation or the indefinite inner product kernel $\hat{\Lambda}\mathbb{I}_{\hat{q}\hat{s}}\hat{\Lambda}^{T}$ as noted above in \eqref{eqn:lambhat}. Plugging this estimate of latent vectors into the least squares objective function \eqref{eqn:LSOrdpg1pls} above, we get:
\begin{eqnarray*}\label{eqn:LSOrdpg1pls1}
	\mathcal{L}(\bgamma;\mathbf{X},A)&=&\sum_{i<j}\left(Y_{ij}(\bgamma)- (\mathcal{F}Y(\bgamma))_{ij}\right)^{2}\nonumber\\
	&=&\sum_{i<j}\left(A_{ij}-x^{T}_{ij}\bgamma - \biggl(\mathcal{F}(A-\sum_{l=1}^{p}\mathbf{X}_{l}\gamma_{l})\biggr)_{ij}\right)^{2},
\end{eqnarray*}
where we recall that $\mathbf{X}_{l}=[x_{l,ij}]$ denotes the $n \times n$ matrix of $l$th edge covariate.
Next, we note that the operator $\mathcal{F}$ is not linear and as a result, a closed-form solution for the linear coefficient vector $\bgamma$ cannot be derived. Clearly, if $\mathcal{F}$ was linear, the final term within brackets in 
the equation above would simplify to $\mathcal{F}A-\sum_{l=1}^{p}\mathcal{F} \mathbf{X}_{l}\gamma_{l}$ and thus lead to estimation of $\bgamma$ through minimization of the least squares error in:
\begin{eqnarray*}
	\tilde A-\widetilde{{\mathcal{F}}A} = (\tilde{\mathbf{X}}- \widetilde{\mathcal{F}\mathbf{X}})\bgamma+ \tilde E,
\end{eqnarray*}
where $\tilde{A}:=\vecc{A}$;  
$\tilde{\mathbf{X}}:=[\vecc{\mathbf{X}_{1}}, \vecc{\mathbf{X}_{2}}, \ldots ,\vecc{\mathbf{X}_{p}}] \in \mathbb{R}^{n^{2} \times p}$;
$\widetilde{{\mathcal{F}}A}:= \vecc\{{\mathcal{F}A}\}$; \\ 
$\widetilde{\mathcal{F}\mathbf{X}}:=[\vecc\{{\mathcal{F}\mathbf{X}_{1}}\},  \vecc\{{\mathcal{F}\mathbf{X}_{2}}\},  \ldots, \vecc\{{\mathcal{F}\mathbf{X}_{p}}\}] \in \mathbb{R}^{n^{2} \times p}$;
$E=(e_{ij})$ denotes the $n \times n$ error matrix and $\tilde{E}:=\vecc\{E\}$;
implying the PLS estimator of $\bgamma$ to be
\begin{equation}\label{eqn:gamhatpls}
	\{(\tilde{\mathbf{X}}- \widetilde{\mathcal{F}\mathbf{X}})^{T}(\tilde{\mathbf{X}}- \widetilde{\mathcal{F}\mathbf{X}})\}^{-1} (\tilde{\mathbf{X}}- \widetilde{\mathcal{F}\mathbf{X}})^{T}(\tilde A-\widetilde{{\mathcal{F}}A}),
\end{equation}
as in classical semiparametric literature, \cite{Fan2005profile}.
However, due to the lack of linearity of the operator $\mathcal{F},$ \eqref{eqn:gamhatpls} is not a valid estimator of $\bgamma$. In the next subsection, we show how an iterative approach to obtain profile least squares estimates of the model can be derived following the framework outlined above.
\subsection{Iterative profile least squares estimation}\label{sec:iterPLS}
We begin with the least squares objective function, exactly as above, however, with a view to design an iterative approach to profile least squares estimation of the unknown model parameters. 
As noted above, for a given $\bgamma$,
we may estimate $\balpha$ or $\Lambda$ via spectral embedding on 
$Y=[Y_{ij}]$ following \cite{Rubin2022}.  
Thus, we proceed by replacing the unknown kernel $\balpha^{T}_{i}\mathbb{I}_{qs}\balpha_{j}$ in \eqref{eqn:LSOrdpg}, with its estimate, 
$(\hat{\Lambda}\mathbb{I}_{\hat{q}\hat{s}}\hat{\Lambda}^{T})_{ij}$, which leads to:
\begin{eqnarray}\label{eqn:iter1}
	\mathcal{L}(\bgamma;\mathbf{X},A)&=&\sum_{i<j}\left(A_{ij}-x^{T}_{ij}\bgamma -(\hat{\Lambda}\mathbb{I}_{\hat{q}\hat{s}}\hat{\Lambda}^{T})_{ij} \right)^{2}\nonumber\\
	&=&\sum_{i<j}\left(Y^{\prime}_{ij}(\hat{\Lambda})-x^{T}_{ij}\bgamma \right)^{2},
\end{eqnarray}
where for a given $\Lambda$, we defined $Y^{\prime}_{ij}(\Lambda):=A_{ij}-({\Lambda}\mathbb{I}_{qs}{\Lambda}^{T})_{ij}$.
Then, it follows that
\begin{equation}\label{eqn:gamhat}
	\hat{\bgamma}(\hat{\Lambda})=(\tilde{\mathbf{X}}^{T}\tilde{\mathbf{X}})^{-1}\tilde{\mathbf{X}}^{T}\tilde{\mathbf{Y}^{\prime}},
\end{equation}
where $\tilde{\mathbf{Y}^{\prime}}=\vecc\{\mathbf{Y}^{\prime}\}$, stacks the transposed rows of  the $n \times n$ matrix $\mathbf{Y}^{\prime}=(Y^{\prime}_{ij})$ together resulting in an $n^2 \times 1$ vector, and $\tilde{\mathbf{X}}:=[\vecc{\mathbf{X}_{1}}, \vecc{\mathbf{X}_{2}}, \ldots ,\vecc{\mathbf{X}_{p}}] \in \mathbb{R}^{n^{2} \times p}$.
This implies an iterative approach where starting with an initial value of $\gamma$ denoted as $\hat{\gamma}^{(m=0)}$, we proceed iteratively by estimating
$\hat{\Lambda}^{(m)}=\hat{\Lambda}(\hat{\gamma}^{(m)})$ via spectral decomposition of $Y(\hat{\gamma}^{(m)})$ (as described above \eqref{eqn:lambhat})
and subsequently update $\hat{\gamma}^{(m)}\rightarrow \hat{\gamma}^{(m+1)}(\hat{\Lambda}^{(m)})$ using \eqref{eqn:gamhat}.
Exact details of the resulting iterative profiles least squares algorithm are given in \cref{alg:PLS}. Let $\hat{\bgamma}:=\hat{\bgamma}^{(M)}$, $\hat{\Lambda}:=\hat{\Lambda}^{(M)}$, or $\hat{\balpha}_{i}:=\hat{\balpha}^{(M)}_{i}, i \in [n]$,
denote the final estimates obtained via this iterative profile least squares approach in $M$ iterations, with $\hat{q}:=\hat{q}_{M}, \hat{s}:=\hat{s}_{M}$ and $\hat{d}:=\hat{d}_{M}$.

\begin{algorithm}[h!]
	\SetAlgoLined
	\DontPrintSemicolon 
	\KwInput{Adjacency $A\in\{0,1\}^{n\times n}$; edge covariates $X_{l}=[x_{l,ij}]\in \mathbb{R}^{n\times n}$, $l\in[p]$}
	\KwOutput{$\{\hat{\bgamma}\in \mathbb{R}^{p\times 1},\hat{\Lambda}=[\hat{\balpha}_{1},\ldots,\hat{\balpha}_{n}]^{T}\in \mathbb{R}^{n\times \hat{d}}\}$}
	{\vspace{.2\baselineskip}
		At step $m=0$, initialize $\hat{\bgamma}^{(m)}=\hat{\bgamma}^{(0)}$  \;}
{\vspace{.2\baselineskip}
At step $m$, compute $Y_{m}=[Y_{ij,m}]$, where $Y_{ij,m}=A_{ij}-x^{T}_{ij}\hat{\bgamma}^{(m)}$    \;}
{\vspace{.2\baselineskip}
Define $\hat{\Lambda}^{(m)}=\hat{U}_{m}|\hat{S}_{m}|^{1/2}$ with $\hat{S}_{m}$ as the diagonal matrix with the $\hat{d}_{m}$ (\cite{ZhuGhodsi}) largest eigenvalues of $|Y_{m}|=({Y}^{T}_{m}{Y}_{m})^{1/2}$ in magnitude on its diagonal, arranged in decreasing order (based on their actual, signed, value), and $\hat{U}_{m}$ as the $n\times \hat{d}_{m}$ matrix with columns as the corresponding orthonormal eigenvectors arranged in the same order\;}
{\vspace{.2\baselineskip}
Set $\hat{R}_{ij}^{(m)}=[\hat{\Lambda}^{(m)}\mathbb{I}_{\hat{q}_{m}\hat{s}_{m}}\hat{\Lambda}^{(m)T}]_{ij}$ with $\hat{q}_{m}+\hat{s}_{m}=\hat{d}_{m}$\;}
{\vspace{.2\baselineskip}
Set $\mathbf{Y}^{\prime}_{m}=[Y^{\prime}_{ij,m}]$ where $Y^{\prime}_{ij,m}=A_{ij}-\hat{R}^{(m)}_{ij}$  \;}
{\vspace{.2\baselineskip}
Update $\hat{\gamma}^{(m)}$ to $\hat{\gamma}^{(m+1)}$ where 
$\hat{\gamma}^{(m+1)}:=(\tilde{\mathbf{X}}^{T}\tilde{\mathbf{X}})^{-1}\tilde{\mathbf{X}}^{T}\tilde{\mathbf{Y}^{\prime}}_{m}$
with $\tilde{\mathbf{Y}^{\prime}}_{m}=\vecc\{{\mathbf{Y}^{\prime}}_{m}\}$, and $\tilde{\mathbf{X}}=\begin{bmatrix}\vecc\{\mathbf{X}_{1}\}, & \ldots & ,\vecc\{\mathbf{X}_{p}\}\end{bmatrix}$\;}
{\vspace{.2\baselineskip}
Update $m \rightarrow m+1$ and go to step 2\;}
{\vspace{.2\baselineskip}
Repeat steps 2-7 for $M-1$ iterations until convergence (with a user specified tolerance on the least squares criterion \eqref{eqn:LSOrdpg} or with a specified maximum number of iterations). Let $\hat{\bgamma}=\hat{\bgamma}^{(M)},\hat{\Lambda}=\hat{\Lambda}^{(M)}$ denote these estimates. Set $\hat{q}:=\hat{q}_{M}$, $\hat{s}:=\hat{s}_{M}$, and $\hat{d}:=\hat{d}_{M}$\;}
\caption{\label{alg:PLS} Profile least squares estimator.}
\end{algorithm}

The estimated parameters 
are subsequently used to construct the $n \times n$ edge probability matrix estimate $\hat{P}=[\hat{P}_{ij}]$ as follows.
We first fit a Gaussian Mixture Model (GMM) on the estimated residual node positions in $\hat{\Lambda}$ to obtain $\hat{K}$ cluster means $\hat{\bmu}_{a}=[\hat{\mu}_{a1},\ldots,\hat{\mu}_{a\hat{d}}]^{T}, a\in[\hat{K}]$ and corresponding node classification vector $\hat{z}=[\hat{z}_{1},\ldots,\hat{z}_{n}]^T, \hat{z}_{i}\in \{1,\ldots,\hat{K}\}$, using which the $\hat{K} \times \hat{K}$ residual matrix $\hat{\Theta}$ follows as $\hat{\Theta}_{ij}=\hat{\bmu}^{T}_{\hat{z}_{i}}\mathbb{I}_{\hat{q}\hat{s}}\hat{\bmu}_{\hat{z}_{j}}$.
Thus, we define $\hat{P}_{ij}=x^{T}_{ij}\hat{\gamma} + \hat{\Theta}_{{i}{j}}.$ The use of GMM for clustering of the estimated latent positions in the case where the residual structure is of a blockmodel type (type I. above) is theoretically justified due to the central limit theorem for adjacency spectral embedding established under this setting (\cite{Rubin2022}). 
Given the widespread applicability of GMMs as the basic model for clustering multivariate data and their flexibility (e.g. \cite{Grun2019}), we tested clustering of estimated latent positions using the same approach when the residual structure is of type II, and found this to lead to satisfactory results in our simulation study (as reported in \cref{sec:sim2}).
For implementation of model based clustering and classification via GMM, the \textit{Mclust} function in statistical software $\textsf{R}$, with $\hat{K}$ selected according to the Bayesian information
criterion (BIC), was used.

In practice, $p$ may not be small, and a convenient choice to initialize the linear coefficient parameter vector $\bgamma$ is as a constant vector, i.e. setting $\hat{\gamma}^{(0)}_{l}=c$ for all $l \in [p]$ where $c$ is a constant.  
To ensure robustness of results, we recommend running the proposed algorithm with different initializations, e.g. varying $c$ over $T$ equispaced values in an interval $[t_{0},t_{1}]$ and subsequently selecting the estimate that leads to the minimum value of the least squares criterion over the chosen set of $T$ initializations. 

\section{Inference via bootstrap}\label{sec:bootstrap}
In this section we show how inference on parameters of our model can be made using the generalized bootstrap for estimating equations \cite{Chatterjee2005generalized}. To achieve this, we first note that
estimating equations for our unknown model parameters, easily follow under the proposed profile least squares approach via \eqref{eqn:LSOrdpg1pls} and \eqref{eqn:iter1}. 
For example, estimating equations for the parameter $\bgamma$ are the set of normal equations given by $\sum_{r=1}^{n^{2}} \mathbf{\tilde{x}_{r}}(\tilde{y}^{\prime}_{r}-\mathbf{\tilde{x}}_{r}^{T} \bgamma)=0,$
where $\tilde{y}^{\prime}_{r}$ denotes the $r$th element of the $n^{2}\times 1$ vector $\vecc\{\mathbf{Y}^{\prime}\}$; and $\mathbf{\tilde{x}}_{r}=[\tilde{x}_{r1},\ldots,\tilde{x}_{rp}]^{T}$ denotes the $r$th (transposed) row of the $n^{2}\times p$ matrix $\tilde{\mathbf{X}}.$
In general for an unknown parameter $\bbeta \in \mathbb{R}^{p}$, estimated by solving equations of the form $\sum_{i=1}^{n}\phi_{ni}(Z_{ni},\bbeta)=0,$
where $\phi_{ni}, 1\leq i\leq n, n\geq 1$ is a triangular sequence of functions taking values in $\mathbb{R}^{p}$ and $\{Z_{ni}\}$ are observable random variables, the generalized bootstrap approach of \cite{Chatterjee2005generalized} shows that the resampling estimator $\hat{\bbeta}^{\ast}$ obtained as the solution to $\sum_{i=1}^{n}w_{ni}\phi_{ni}(Z_{ni},\bbeta)=0,$
has desirable theoretical properties (e.g. distributional consistency) under suitable conditions on the random bootstrap weights $\{w_{ni}, 1\leq i\leq n, n\geq 1\}$. 
Following the results of their work, we describe the bootstrap approach for inference on the unknown model parameters of \eqref{eqn:RDPGcov} below.

Let $\mathbf{W}_{b} = (W_{b1}, W_{b2}, \ldots, W_{bn}) \in \mathbb{R}^{n}$ be a random vector that we shall refer to as resampling (or bootstrap) weights. Thus, to obtain $B$ bootstrap samples we require $\{ \mathbf{W}_{1}, \ldots, \mathbf{W}_{B} \}$, where each $\mathbf{W}_{b}$ is a $n$-dimensional vector. 
Different choices of the distribution of bootstrap weights lead to different bootstrap techniques. For example, the classical or naive bootstrap technique where each $\mathbf{W}_{b} \iid \text{Multinomial}\bigl(n; 1/n, 1/n, \ldots, 1/n  \bigr)$; or the $m$-out-of-$n$ (moon) bootstrap  where each $\mathbf{W}_{b} \iid \text{Multinomial} \bigl(m; 1/n, 1/n, \ldots, 1/n  \bigr)$.
Finally, the bayesian bootstrap employs for each $b \in[B]$ and  $i\in [n]$, $W_{bi} \iid \ \text{Exponential}(\alpha)$, typically with $\alpha = 1$ and $\alpha = O (n^{-1/2})$ as these two choices have performances comparable to the naive bootstrap and the moon bootstrap. Bayesian bootstrap has several theoretical and computational benefits 
and hence in what follows, we shall use this approach. 

Once we have obtained the resampling weights $\{ \mathbf{W}_{1}, \ldots, \mathbf{W}_{B} \}$,
the $b$-th resample version of parameters, denoted as $\hat{\Lambda}^{\ast}_{b}=[\hat{\balpha}^{\ast T}_{1b},\ldots,\hat{\balpha}^{\ast T}_{nb}]^{T}\in \mathbb{R}^{n \times \hat{d}}$ and $\hat{\bgamma}^{\ast}_{b}$ are obtained as described below. First, obtain $\hat{\bgamma}_{b}^{\ast}$
by minimizing the residual sum of squares:
\begin{equation}\label{eqn:bt2}
\mathcal{L}_{b}(\bgamma;\hat{\Lambda},\mathbf{X},A)=\sum_{i<j}W_{bi}W_{bj}\left(Y^{\prime}_{ij}(\hat{\Lambda})-x^{T}_{ij}\bgamma \right)^{2},
\end{equation}
where $Y^{\prime}_{ij}=A_{ij}-(\hat{\Lambda}\mathbb{I}_{\hat{q} \hat{s}}\hat{\Lambda}^{T})_{ij}$, follows from the definition of $Y^{\prime}$ in \cref{sec:iterPLS} and where $\hat{\Lambda}$ denotes the final profile least squares estimate of latent positions as obtained via \cref{alg:PLS}.
Next, the resampling estimates $\hat{\Lambda}^{\ast}_{b}$ for $b\in[B]$ are obtained as follows. First let $\hat{\Lambda}^{w}_{b}=[\hat{\balpha}^{w T}_{1b},\ldots,\hat{\balpha}^{w T}_{nb}]^{T}$ denote the $\hat{d}$-dimensional latent position vectors obtained by minimizing the residual sum of squares:
\begin{equation}\label{eqn:bt1}
\mathcal{L}_{b}(\Lambda;\hat{\gamma}^{\ast}_{b},\mathbf{X},A)=\sum_{i<j}W_{bi}W_{bj}\left(Y_{ij}(\hat{\bgamma}^{\ast}_{b})- \balpha^{T}_{i}\mathbb{I}_{qs}\balpha_{j} \right)^{2},
\end{equation}
where now $Y_{ij}=A_{ij}-x^{T}_{ij}\hat{\bgamma}^{\ast}_{b}$, following the definition of $Y$ in \cref{sec:plsexact}.
Next, we note that \eqref{eqn:bt1} clearly implies that instead of resampling ${\balpha}_{i}$, we shall resample its weighted version $\sqrt{W_{bi}}{\balpha}_{ib}$ that we denoted as $\hat{\balpha}^{w}_{ib}$. Thus, to obtain bootstrap samples of latent positions corresponding to our model, we set $\hat{\balpha}^{\ast}_{ib}:=\hat{\balpha}^{w}_{ib}/\sqrt{W_{bi}}$ and $\hat{\Lambda}^{\ast}_{b}=[\hat{\balpha}^{\ast T}_{1b},\ldots,\hat{\balpha}^{\ast T}_{nb}]^{T}$.

\begin{figure}[tbh]
\centering
\includegraphics[scale=0.5]{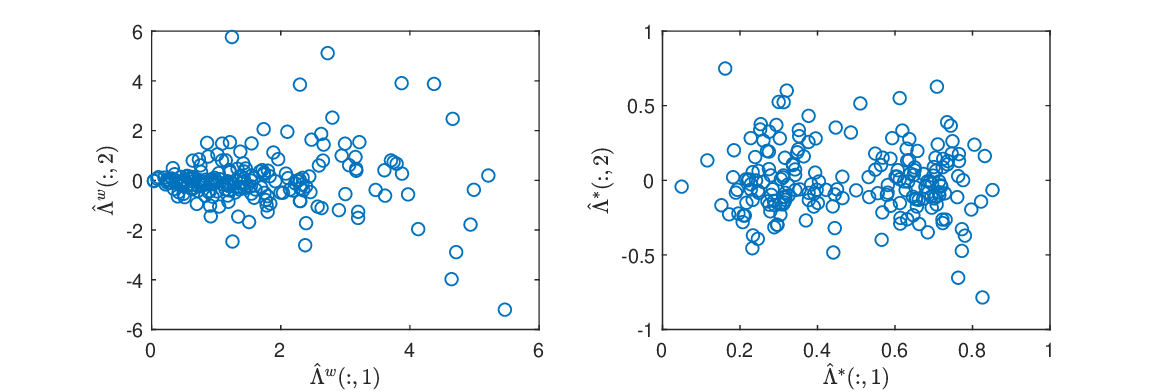}
\caption{Scatter plot of the first two dimensions of the latent positions (a) $\hat{\balpha}^{w}_{ib}$ obtained as the solution to \eqref{eqn:bt1}, and (b) the corresponding de-weighted bootstrap latent positions $\hat{\balpha}^{\ast}_{ib}:=\hat{\balpha}^{w}_{ib}/\sqrt{W_{bi}}$, for the simulation setting (a) described under \cref{sec:SIM1}. Here $n=200$.}
\label{fig:wtdvsdewtd}
\end{figure}

To visualize the importance of de-weighting as done above, \cref{fig:wtdvsdewtd}, displays a scatter plot of the first two dimensions of the latent positions $\hat{\balpha}^{w}_{ib}$ obtained via \eqref{eqn:bt1} (left subplot) and the corresponding de-weighted latent positions $\hat{\balpha}^{\ast}_{ib}$ (right subplot) for one of the simulation examples (described in \cref{sec:SIM1}). As expected, we find that only the de-weighted latent positions (here $d=1$ and $K=2$) are aligned with the ground truth means $\mu_{1}=0.3$ and $\mu_{2}=0.668$ of the two underlying clusters. 

Subsequently, the bootstrap residual matrix $\hat{\Theta}^{\ast}_{b}$ may be obtained via GMM fitting on the bootstrap latent position vectors $\{\hat{\balpha}^{\ast}_{ib}\}_{i=1}^{n}$.  To ensure that 
there is no nuisance variability  in bootstrap samples due to different choice of the number of clusters, we recommend using $\hat{K}$ clusters in each bootstrap replication $b\in[B]$ as used to construct the final residual matrix estimate $\widehat{\Theta}$. Thus, let $\hat{z}^{\ast}_{b}=[\hat{z}^{\ast}_{b1},\ldots,\hat{z}^{\ast}_{bn}]^T, \hat{z}_{bi}\in \{1,\ldots,\hat{K}\}$ denote the bootstrap node classification vector obtained via GMM fitting on the bootstrap latent positions $\{\hat{\balpha}^{\ast}_{ib}\}_{i=1}^{n}$. It is important to note that in general, the cluster labels $a\in\{1,\ldots,\hat{K}\}$ need not be consistent across bootstrap replications, for example, with cluster labeled as 1 in bootstrap replication $b=1$ corresponding to cluster 3 in replication $b=2$. Hence, for consistency and ease of inference via bootstrap, we recommend a re-labeling of estimated residual clusters in each boostrap replication based on a monotonic constraint on their intensities
 i.e. clusters are (re)labeled $1,2,\ldots,\hat{K}$ to correspond to 
$\hat{\Theta}^{\ast}_{1}\geq \widehat{\Theta}^{\ast}_{2}\ldots\geq \widehat{\Theta}^{\ast}_{\hat{K}}$, where $\widehat{\Theta}^{\ast}_{a}=\hat{\mu}^{\ast T}_{a}\mathbb{I}_{\hat{q}\hat{s}}\hat{\mu}^{\ast}_{a}, a\in [\hat{K}]$, with $\hat{\mu}^{\ast}_{a}$ denoting the mean vector of cluster $a$ as estimated via GMM fit on bootstrap latent positions $\{\hat{\balpha}^{\ast}_{ib}\}_{i=1}^{n}$.

\section{Finite sample performance}\label{sec:sims}
To illustrate the performance of our iterative profile least squares estimator, we report results obtained using networks simulated with categorical and/or continuous covariates and with residual structures generated from either a stochastic blockmodel (type I.) or with a general low-rank structure (type II.). Simulations with type I. residual are presented in \cref{sec:SIM1} below, and results for type II. are included in \cref{sec:sim2}. In each case, \cref{alg:PLS} was run with $M=500$ iterations and the final results correspond to the estimate that lead to the minimum value of the least squares criterion starting with $T=20$ equispaced initializations with $\hat{\bgamma}_{l}^{(0)}=c \in [0.15,2]$, $l\in[p]$.

\subsection{Residual structure of Type I.}\label{sec:SIM1}
We generate adjacency $A=[A_{ij}]$ with $A_{ij}, i<j$ sampled as independent Bernoulli trials with success probabilities $P_{ij}$ of the form:
\begin{equation}\label{Pijours}
P_{ij}=x^{T}_{ij}\bgamma + \Theta_{z_{i}z_{j}},
\end{equation}
where $\bgamma=[\gamma_{1},\ldots,\gamma_{p}]^{T}$ denotes the coefficient vector for $p$ pairwise (or edge) covariates in $x_{ij}=[x_{1,ij},\ldots,x_{p,ij}]^{T}$; and with the residual structure arising from a block constant matrix $\Theta_{z_{i}z_{j}}$ where $z=[z_{1},\ldots,z_{n}]^{T}$ and each $z_{i}\in [K]$ denotes the class label to which node $i$ belongs. For $i<j$, set $A_{ji}=A_{ij}$ and $A_{ii}=0$. Following \cite{Muetal}, we chose $K=2$ and  the residual matrix to be of the form 
$\Theta=\begin{bmatrix} \mu_{1}^2 &  \mu_{1} \mu_{2} \\
\mu_{1} \mu_{2} & \mu^{2}_{2}\end{bmatrix},$
for $0<\mu_{1}<\mu_{2}<1$.
This corresponds to a rank one ($d=1$) GRDPG, with latent positions 
$[\mu_{1}, \mu_{2}]^{T}$ 
where for node $i$, $\alpha_{i}=\mu_{1}$ if $z_{i}=1$ and $\alpha_{i}=\mu_{2}$ if $z_{i}=2$. Clearly, $\Theta=[\mu_{1}, \mu_{2}]^{T}[\mu_{1}, \mu_{2}]$.
For convenience, we define $\mu:=[\mu^{2}_{1},\mu^{2}_{2}, \mu_{12}]^{T},$ containing the diagonal and the unique off-diagonal element $\mu_{12}=\mu_{1}\mu_{2}$ of the residual matrix $\Theta$. Then setting $\mu_{1}=0.3$ and $\mu_{2}=0.668$ as in \cite{Muetal}, we have $\mu=[0.09,0.446224,0.2004]^{T}$.

Next, we generated covariates to correspond to three different settings: (a) $p=1$ where for each node $i$, we generate a binary categorical covariate $x_{i}\in \{0,1\}$ from a Bernoulli($0.5$) and set coefficient $\gamma_{1}=0.4$; (b) $p=1$ where for each node $i$, we generate a continuous covariate $x_{i}\in \mathbb{R}$ from a Normal$(0.2,0.25)$ and set coefficient $\gamma_{1}=0.4$, and (c) $p=2$ where for each node $i$, we generate $x_{1i}\in \{0,1\}$ from a Bernoulli($0.5$) and $x_{2i}\in \mathbb{R}$ from a Normal$(0.2,0.25)$ and set $\gamma_{1}=0.4$ and $\gamma_{2}=0.1$.
The first setting (a) above, where an adjacency with a single categorical covariate is observed has been considered in \cite{Muetal}, where specifically, edge probabilities $P_{ij}$ are of the form $P_{ij}=\beta \mathbb{I}\{x_{i}=x_{j}\} + B_{z_{i}z_{j}},$ where $B$ represents a SBM. We note that in the case of a single binary covariate $x_{i}\in \{0,1\}$, with corresponding edge covariate $x_{ij}:=|x_{i}-x_{j}|$, our model in \eqref{Pijours} can alternatively be expressed as $P_{ij}=\gamma |x_{i}-x_{j}|+ \Theta_{z_{i}z_{j}}=-\gamma \mathbb{I}\{x_{i}=x_{j}\}+ (\gamma+ \Theta_{z_{i}z_{j}})$,
showing how our model in 
this special case reduces to the model of \cite{Muetal} with $\beta=-\gamma$ and $B_{z_{i}z_{j}}=\gamma+ \Theta_{z_{i}z_{j}}$. We compare the performance of our PLS estimator with that of \cite{Muetal}, in this case below.

\begin{figure*}
\centering
\includegraphics[width=1\textwidth,scale=0.65]{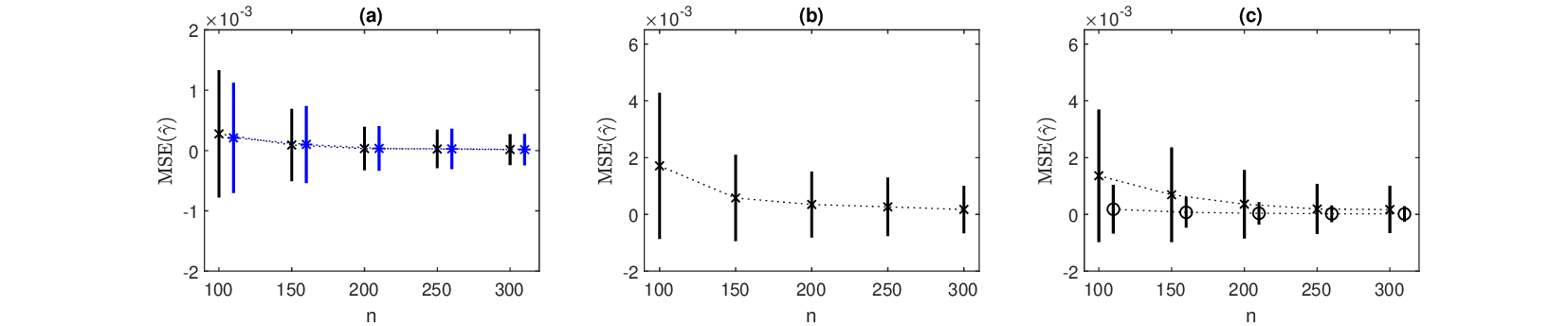}
\caption{Mean squared errors of $\hat{\gamma}_{l}, l\in [p]$ estimated using the proposed profile least squares approach (`x',`$\circ$') and of estimates obtained using the method of  \cite{Muetal} (`$\ast$' ) using networks of size $n \in\{100,150,200,250,300\}$, in each case averaged over $250$ replications, where (a) $\gamma=0.4$ and covariates $x_{i}\sim \text{Bernoulli}(0.5), i\in [n]$ independently, (b)  $\gamma=0.4$ and covariates $x_{i}\sim \text{Normal}(0.2,0.25), i\in[n]$ independently, and (c)  $\gamma=[0.4,0.1]^{T}$ and covariates $x_{1i}\sim \text{Bernoulli}(0.5),$ and $x_{2i}\sim \text{Normal}(0.2,0.25), i\in [n]$,  each independently.}
\label{fig:MSEa}
\end{figure*}

\begin{figure*}
\centering
\includegraphics[width=1\textwidth,scale=0.68]{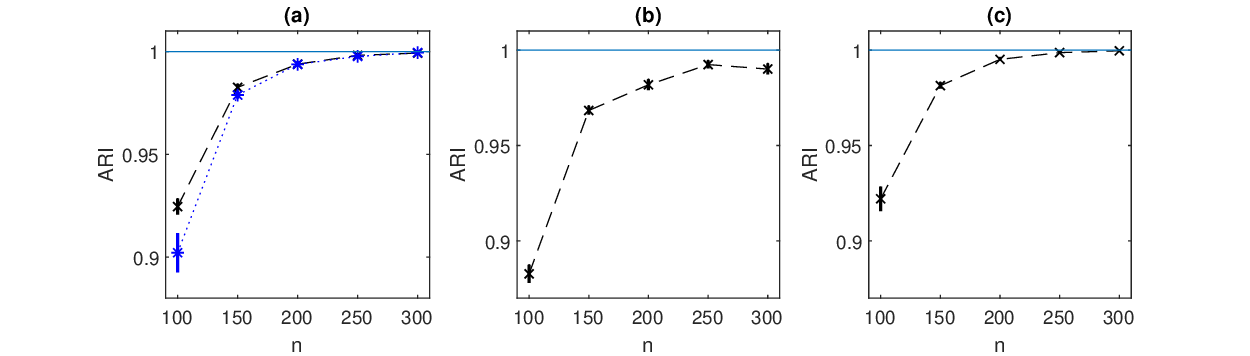}
\caption{Mean Adjusted Rand Index (ARI) based on latent positions estimated using the proposed profile least squares approach (`x' ) and from the method of \cite{Muetal} (`$\ast$') using networks of size $n \in\{100,150,200,250,300\}$  in each case averaged over $250$ samples, where subplots (a)-(c) correspond to the three covariate settings under type I residual as described in the text.} 
\label{fig:ARIb}
\end{figure*}

\cref{fig:MSEa} and \cref{fig:ARIb} display mean squared errors (MSE) of the estimated covariate coefficient vector $\hat{\bgamma}$ and the mean adjusted rand index (ARI) based on clusters identified using the estimated latent positions $[\hat{\balpha}_{i}]_{i \in n}$, respectively, as a function of the size of the network $n \in\{100,150,200,250,300\}$, for the three covariate settings (a)-(c) described above, with estimates obtained using the proposed iterative profile least squares algorithm. As the methodology of \cite{Muetal} applies only in covariate setting (a), results obtained from their algorithm are also included in the corresponding subplots (a). From \cref{fig:MSEa}(a) and \cref{fig:ARIb}(a), it is clear that in the case of a single binary categorical covariate, estimates from our method are comparable to those from \cite{Muetal} for all $n\geq 150$; for the smallest $n=100$, our method leads to a marginally higher ARI score.
In the remaining two settings ((b) and (c)), \cite{Muetal} does not apply and the subplots show how MSE($\hat{\gamma}_{l}$), $l\in[p]$ decays and the mean ARI increases with increase in the network size $n$. 

Note that the mean ARI in the case of the single continuous covariate as displayed in \cref{fig:ARIb}(b) does not approach the maximum value of $1$ as $n$ approaches $300$. We found that this was due to variability in the number of clusters $\hat{K}$ detected via \textit{MClust} in our default implementation of the GMM clustering of estimated latent positions.
Specifically, it was observed that in a few replications ($11$ out of $250$), \textit{MClust} detected $\hat{K}=3$ clusters (instead of $\hat{K}=2$ as in the majority of replications),
which led to an ARI less than $1$. 
This, relatively less stable detection of the number of clusters $K$ in this case is not surprising since latent positions are estimated via spectral embedding of the residual (non-binary) matrix $A_{ij}-x^{T}_{ij}\bgamma$ which is relatively less discrete when only a single continuous covariate is observed in comparison to cases (a) and (c) where a categorical covariate is observed. 
In each of these replications, a scatter plot of the first two dimensions of the estimated latent position vector clearly indicates two clusters. Thus, we re-ran the final clustering step for these replications by inputting the number of clusters as $\hat{K}=2$ in \textit{MClust}. This significantly reduced the average uncertainty of the estimated classification (from $0.085$ to $10^{-6}$) indicating that $\hat{K}=2$ is a more robust choice for clustering in these replications as well. Further, this increased the overall mean ARI to $1$ for $n=250$ and $n=300$. Thus, in practice, where only a single continuous covariate is observed and interest lies in the residual structure, we recommend following the procedure above i.e. using a scatter plot to visually detect a potential alternative to the $\hat{K}$ selected via default implementation of \textit{MClust} along with the average uncertainty of the clustering (outputted by the \textit{MClust} function) as an indicator of a robust choice of the number of clusters.

\begin{figure}[tbh]
\centering
\includegraphics[scale=0.54]{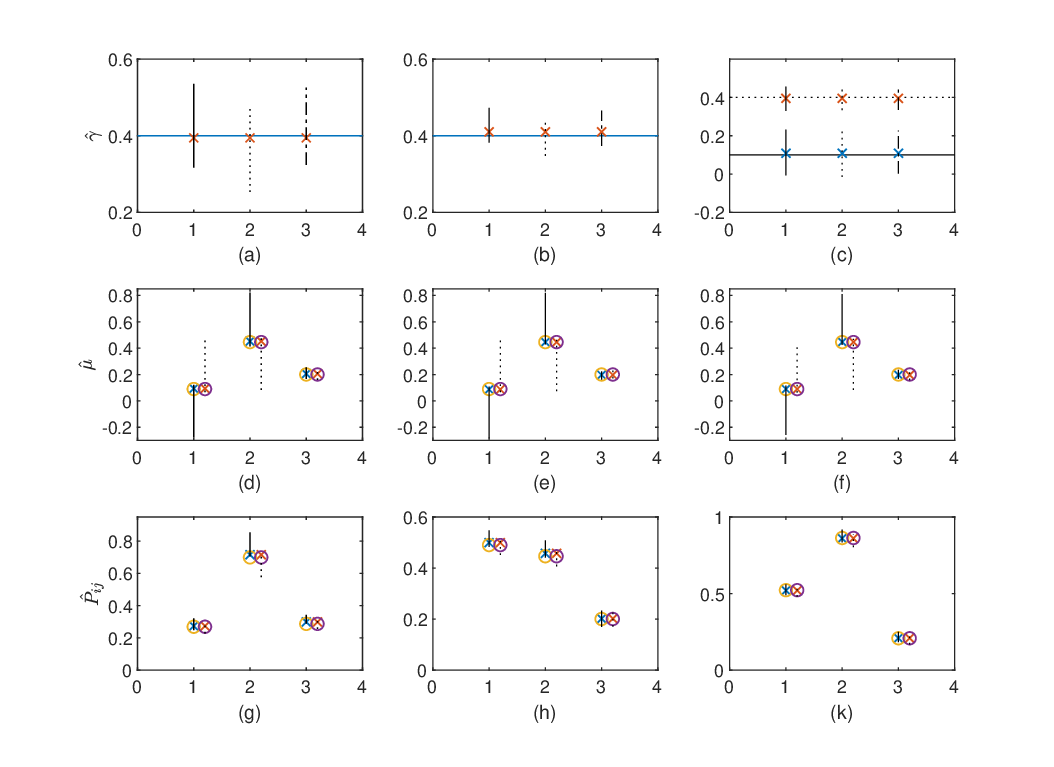}
\caption{True and estimated parameters ($\gamma, \mu=[\mu^{2}_{1},\mu^{2}_{2},\mu_{12}]^{T}$) and resulting edge probabilities $P_{ij}$ in rows 1-3 for covariate settings (a)-(c) in columns 1-3, respectively, estimated from a network of size $n=200$. The vertical bars display the corresponding $95\%$ bootstrap confidence intervals as follows. First row: covariate coefficient $\hat{\gamma}_{l}, l\in [p]$ (`x') vs true $\gamma_{l}$ with the basic (solid), percentile (dotted) and normal with bias correction (dot-dash) confidence intervals; second row: true (`$\circ$') and estimated (`x') components of the residual vector $\mu$, with the basic (solid) and percentile (dotted) confidence intervals; third row: true (`$\circ$') and estimated (`x') edge probabilities for three node pairs $(i,j)$ mapped to block pairs $\{(1,1), (2,2), (1,2)\}$ with the basic (solid) and percentile (dotted) confidence intervals.}
\label{fig:btsim}
\end{figure}

\cref{fig:btsim} displays estimates of the two components ($\hat{\gamma}$ and $\hat{\mu}$) of our model in \eqref{Pijours} and the  resulting edge probabilities $\hat{P}_{\hat{z}_{i}\hat{z}_{j}}$ for $(\hat{z}_{i},\hat{z}_{j})=(1,1), (2,2)$ and $(1,2)$ estimated via the proposed method along with corresponding $95\%$ bootstrap confidence intervals using $B=1000$ bootstrap replications (bayesian bootstrap with weights $\alpha=n^{-1/2}$), for a randomly generated network of size $n=200$ under the three covariate settings in subplots (a)-(c). 
Clearly, the resulting bootstrap confidence intervals contain the true value of the parameter in each setting, confirming the suitability of the proposed bootstrap methodology for inference on parameters of our model.
The confidence intervals clearly vary depending on their construction, as is commonly observed in the bootstrap literature, e.g. \cite{Davison1997}.

\subsection{Residual structure of Type II.}\label{sec:sim2}
Here, adjacency matrices $A=[A_{ij}]$ were generated with $A_{ij}, i<j$ sampled as independent Bernoulli trials with success probabilities $P_{ij}$ given by \eqref{Pijours},
where as before, $\gamma=[\gamma_{1},\ldots,\gamma_{p}]^{T}$ denotes the coefficient vector for $p$ pairwise (or edge) covariates in $x_{ij}=[x_{1,ij},\ldots, x_{p,ij}]^{T}$; and the residual structure is of a block constant form $\check{\Theta}_{z_{i}z_{j}} \in \mathbb{R}^{K \times K}$ where $z=[z_{1},\ldots,z_{n}]^{T}$ and each $z_{i}\in [K]$ denotes the cluster index to which node $i$ belongs. 

We chose $d=2$-dimensional latent position vectors given by
$\bmu_{1}=[\sqrt{0.6},\sqrt{0.3}]^{T}$ and $\bmu_{2}=[\sqrt{0.2},\sqrt{0.4}]^{T}$ with $q=1$ and $s=-1$, implying 
that the $2$-block residual kernel is given by 
\begin{equation}\label{eqn:resid2}
	\check{\Theta}=\begin{bmatrix}  
		0.3 & 0\\
		0 & -0.2
	\end{bmatrix},
\end{equation}
since $\check{\Theta}_{11}=\mu^{2}_{11}-\mu^{2}_{12}$; $\check{\Theta}_{22}=\mu^{2}_{21}-\mu^{2}_{22}$ and $\check{\Theta}_{12}=\mu_{11}\mu_{21}-\mu_{12}\mu_{22}=\check{\Theta}_{21}$ as noted in the example presented in \cref{sec:model}.
We assume that nodes are assigned to one of the two blocks with probabilities $\pi=(1/3, 2/3)$, and thus an imbalanced design.
Recall that to generate valid edge probabilities, chosen parameters must satisfy constraints as stated in \cref{sec:model}. Thus, we shall generate covariates taking the specific residual structure given by \eqref{eqn:resid2} into consideration. We consider the three covariate settings in (a)-(c) below.
\begin{enumerate}
	\item[(a)] Consider the case of a single binary covariate $x_{i}\in \{0,1\}$, where as above, we define $x_{ij}=|x_{i}-x_{j}|$. Then, we note that residual structure of type II cannot arise in this setting. This is because the minimum and maximum possible values of the edge covariate in this case are $0$ and $1$, respectively, implying that
	\begin{equation}
		P_{ij}=\begin{cases} \check{\Theta}_{z_{i}z_{j}} +\gamma \mbox{ if } x_{ij}=1\\
			\check{\Theta}_{z_{i}z_{j}}  \mbox{ if } x_{ij}=0.
		\end{cases}
	\end{equation}
	Since $P_{ij}$ reduces to $\check{\Theta}_{z_{i}z_{j}}$ for all $(i,j)$ where $x_{ij}=0$, we must have $0\leq \check{\Theta}_{z_{i}z_{j}}\leq 1$ to ensure valid edge probabilities in $[0,1]$. In other words, both the conditions of \eqref{eqn:const2} shall only be satisfied when the residual matrix is of type I. i.e. for $0\leq \check{\Theta}_{ab}\leq 1$ for all $(a,b)\in [K]\times [K]$. Hence we do not need to consider this setting.
	\item[(b)] In the case of a single $p=1$ continuous covariate, 
	we directly generated a symmetric edge covariate matrix $[x_{ij}]$ in a $3 \times 3$ block-form as follows. 
	Specifically $x_{ij}$ in blocks $(1,1), (2,2)$ and $(1,2)$ were generated as i.i.d. normal random variables with mean and standard deviation parameters set to 
	$\mu_{x,11}=0.3$, $\mu_{x,22}=0.9$, $\mu_{x,12}=0.3$ and $\sigma_{x}=1/16=0.0625,$ respectively. Edge covariate values in the remaining blocks i.e.~$(1,3), (2,3)$ and $(3,3)$
	were set to $0.4$.
	\cref{fig:simdataII} displays the sampled edge covariate matrix along with the residual structure of \eqref{eqn:resid2} and the corresponding edge probability matrix implied via our model $P_{ij}=\gamma x_{ij} + \check{\Theta}_{z_{i}z_{j}},$ with $n=150$ and $\gamma=0.7$.
	\begin{figure}
		\centering
		\includegraphics[scale=0.45]{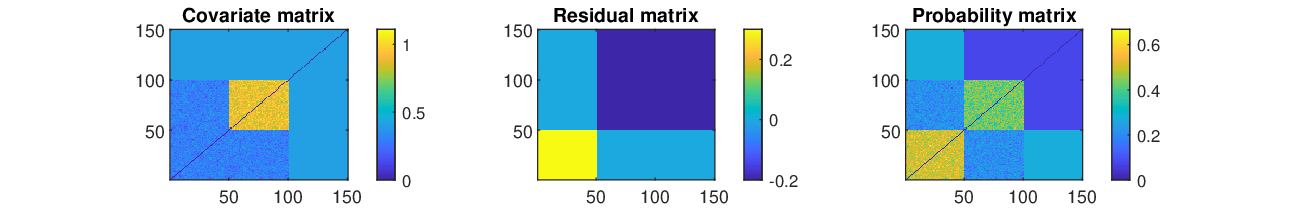}
		\caption{Ground truth data example with a single continuous covariate ($p=1$) requiring type II residual structure. \textit{Left to right}: Edge covariate matrix $[x_{ij}]$;  the residual matrix implied by $\check{\Theta}$ with nodes assigned to blocks with probabilities $\pi=(1/3,2/3)$; the edge probability matrix $P_{ij}=0.7 x_{ij} + \check{\Theta}_{z_{i}z_{j}}$. }
		\label{fig:simdataII}
	\end{figure}

\begin{figure*}
	\centering
	\includegraphics[scale=0.65]{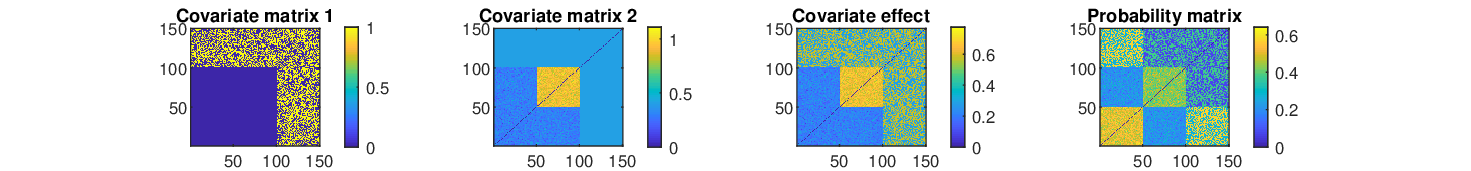}
	\caption{Ground truth data example with a discrete and a continuous covariate ($p=2$) requiring type II residual structure. \textit{Left to right}: discrete edge covariate $x_{1,ij}$; continuous edge covariate $x_{2,ij}$;
		total covariate effect: $0.3 x_{1,ij}+0.7x_{2,ij}$; the edge probability matrix $P_{ij}=0.3 x_{1,ij} + 0.7 x_{2,ij}+\check{\Theta}_{z_{i}z_{j}}$.}
\label{fig:simdataIIc}
\end{figure*}
	\item[(c)] 
	Consider the case of $p=2$ covariates, with categorical $x_{1i}\in \{0,1\}$ and continuous valued $x_{2i}$ such that edge covariates $x_{2,ij}$ are generated directly as described above in (b). To generate the discrete binary covariate, we again proceeded by generating a symmetric edge covariate matrix directly and following the $3 \times 3$ block-form as used in (b). We set $x_{1,ij}=0$ for all node pairs in blocks $(1,1), (2,2), (1,2), (2,1)$, and subsequently generated $x_{1,ij} \sim $Ber(0.5) for node pairs in the remaining blocks $(1,3), (2,3)$ and $(3,3)$. Here we set $\gamma_{1}=0.3$ and $\gamma_{2}=0.7$ (as in (b)). \cref{fig:simdataIIc} displays the two edge covariate matrices along with the total covariate effect $0.3 x_{1,ij}+0.7x_{2,ij}$ and the corresponding edge probability matrix $P_{ij}$ with $n=150$.
\end{enumerate}

\begin{figure}
\centering
\includegraphics[scale=0.55]{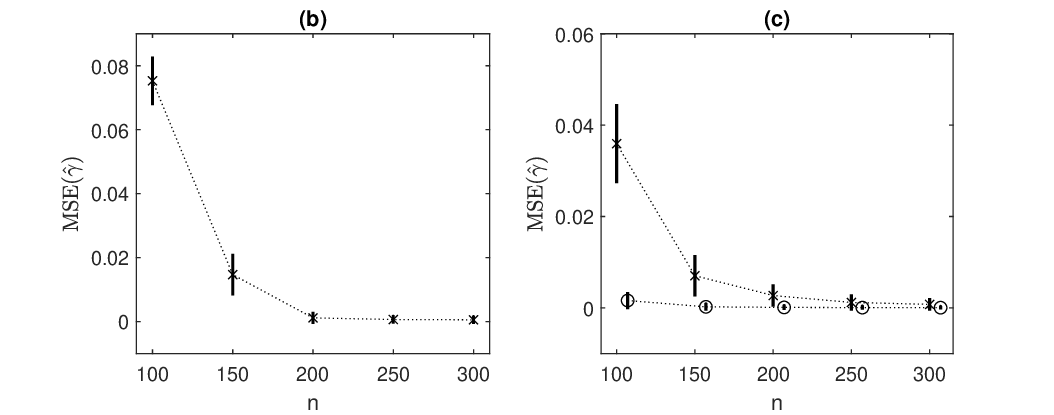}
\bigbreak
\includegraphics[scale=0.65]{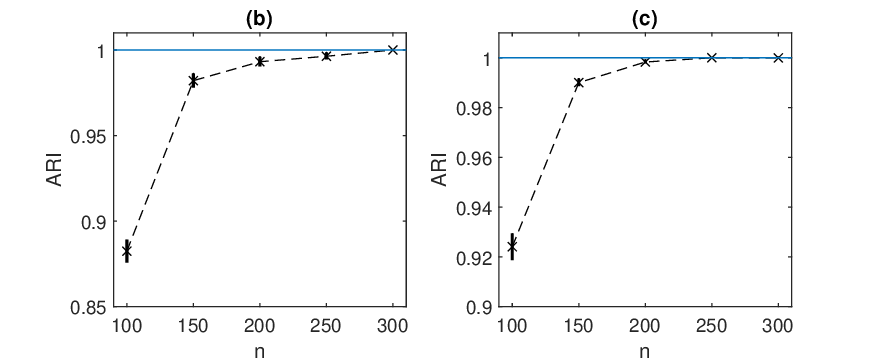}
\caption{\textit{First row}: Mean squared errors of $\hat{\gamma}_{l}, l\in [p]$ estimated using the proposed profile least squares approach using networks of size $n \in\{100,150,200,250,300\}$, in each case averaged over $250$ replications, under covariate setting (b) $\gamma=0.7$ with a single continuous-valued covariate $x_{i}, i\in [n]$, and (c)  $\gamma=[0.3,0.7]^{T}$ and covariates $x_{1i}\sim \text{Bernoulli}(0.5),$ and $x_{2,ij}$ as described in the text. \textit{Second row}: the corresponding mean ARI scores. Vertical bars display the corresponding standard errors.}
\label{fig:MSEtype2}
\end{figure}
\begin{figure}
\centering
\includegraphics[scale=0.6]{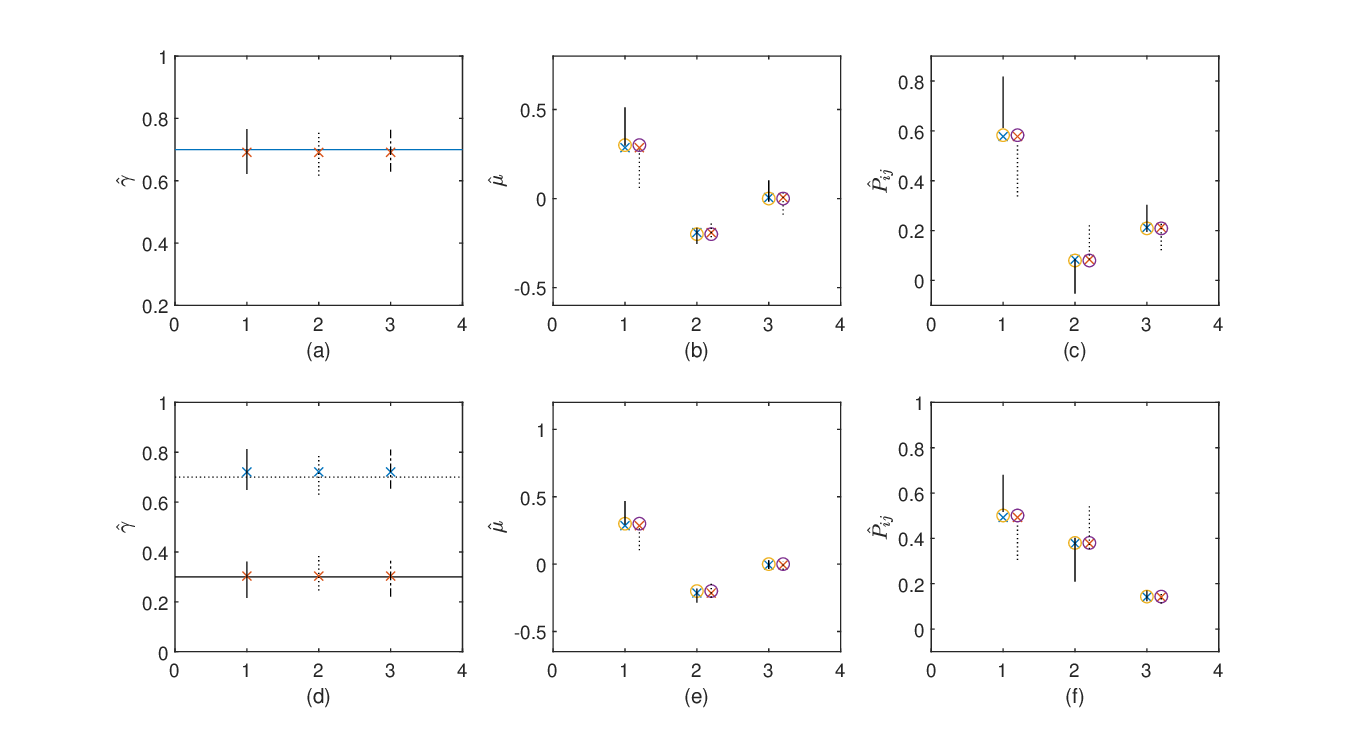}
\caption{ Estimates of the covariate coefficient vector $\gamma$ (\textit{left}), residual structure intensities $\mu$ (\textit{middle}) and implied edge probabilities $P_{ij}$ for three pairs of nodes $(i,j) | (\hat{z}_{i},\hat{z}_{j})=(1,1), (2,2), (1,2)$ (\textit{right}), with bootstrap confidence intervals for the corresponding true values (lines and labels as in \cref{fig:btsim}). \textit{First row}: $p=1$ case with a single continuous covariate;  \textit{second row}: $p=2$ case with a binary categorical and a continuous covariate (as described in the text).}
\label{fig:btsim2}
\end{figure}

MSE($\hat{\bgamma}$) and mean ARI comparisons for networks of sizes $n\in\{100,150,200,250,300\}$ with covariates generated as described in settings (b) and (c) are displayed in \cref{fig:MSEtype2}. Clearly, our method leads to satisfactory results with the MSE($\hat{\bgamma}$) almost equal to $0$ for $n\geq 200$.
\cref{fig:btsim2} displays estimates of $\hat{\gamma}$, $\hat{\mu}$ and the corresponding edge probabilities $\hat{P}_{\hat{z}_{i}\hat{z}_{j}}$ for $(\hat{z}_{i},\hat{z}_{j})=(1,1), (2,2)$ and $(1,2)$ from the proposed method along with corresponding $95\%$ bootstrap confidence intervals based on $B=1000$ replications for a randomly generated network of size $n=200$ under the two covariate settings (b), (c). Clearly, bootstrap confidence intervals in each case contain the corresponding true unknown value of the model parameter (and the edge probabilities they imply), confirming the validity of the proposed bootstrap approach for inference.

\section{Network data examples}
We applied our methodology and inference to four real network datasets which observe covariates of different types. 
Following the initialization strategy described in \cref{sec:iterPLS}, for all data illustrations included below, we obtained results from our algorithm with $T=20$ equispaced initial values $\hat{\gamma}^{(0)}_{l}=c \in [0.15,2], \forall l\in [p]$.
In all data implementations except the alliance network in year 1995, our algorithm converged to the same estimate of the model 
irrespective of the initial values.
For the alliance network in 1995, we obtained three different estimates over these $T=20$ initializations and thus report estimates which correspond to the lowest value of the least squares objective function. 
For inference on model parameters, the proposed bayesian bootstrap procedure with weights $\alpha=n^{-1/2}$ and $B=999$, 
is employed.
To visualize the contribution of the observed covariates to the overall network topology, we display the estimated covariate effect matrix $\hat{C}_{ij}:=x_{ij}^{T}\hat{\bgamma}$, the residual structure $\hat{\Theta}_{ij}$ and the resulting edge probability matrix $\hat{P}_{ij}$ from our proposed method. For a smooth visualization, we display all three matrix estimates with nodes permuted to correspond to the relabeling of clusters in the latent structure such that $\hat{\Theta}_{11}\geq \hat{\Theta}_{22}\ldots\geq \hat{\Theta}_{\hat{K}\hat{K}}$.
Likewise, we display the edge probability matrix estimated via the GRDPG approach (\cite{Rubin2022}) with its rows and columns permuted such that edge probabilities on the diagonal appear in a descending order. 

\subsection{Tree network}
\begin{figure}
\centering
\includegraphics[scale=0.4]{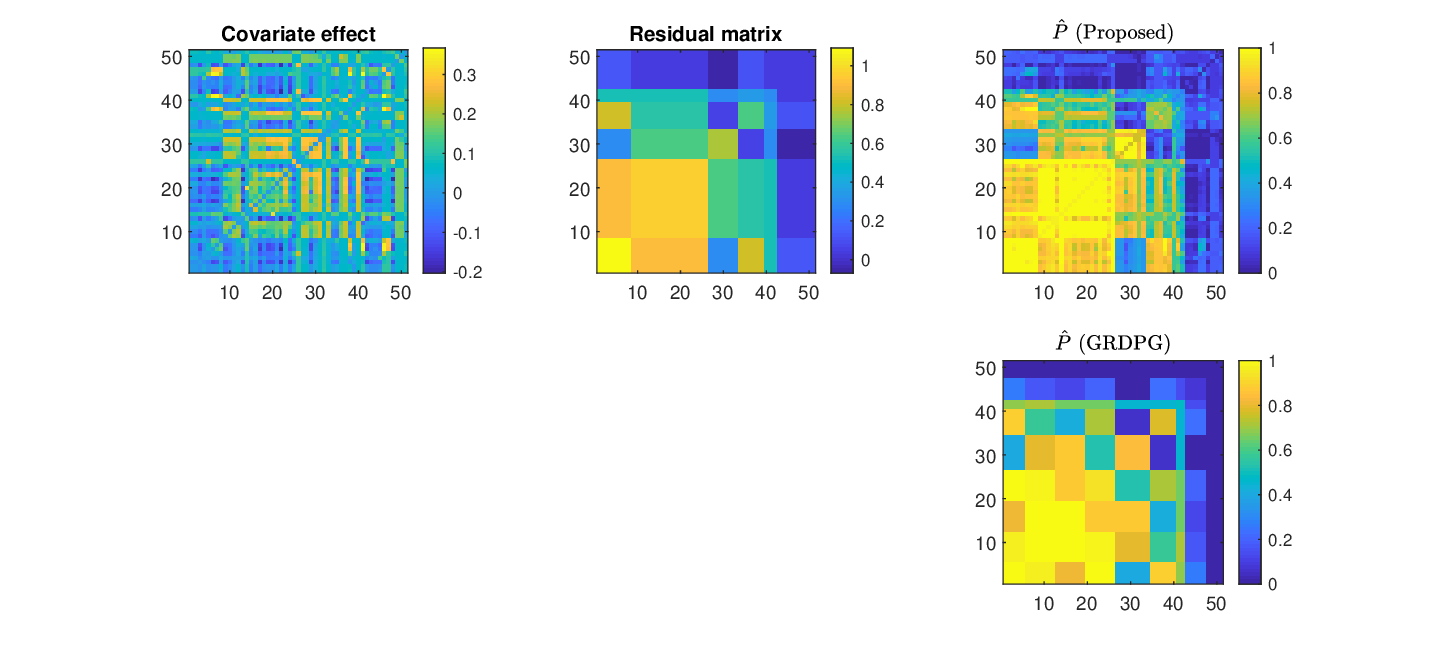}
\caption{\textit{First row}: Estimated covariate effect $[\hat{C}_{ij}]$, residual structure $[\hat{\Theta}_{ij}]$, and the corresponding edge probability matrix $\hat{P}=[\hat{P}_{ij}]_{n \times n}$ via the proposed PLS; \textit{Second row}: $\hat{P}$ from ASE(A) under GRDPG, for the tree network.}\label{fig:TreePLS}
\end{figure}

This is a network on $n=51$ trees where an edge exists between two trees if they share at least one common fungal parasite, \cite{Latouche15}, \cite{Vacher2008}, \cite{mariadassou2010}. The genetic, taxonomic and geographic distances between the tree species are observed and hence $p=3$ and we directly observe edge covariates $x_{l,ij}, l=1,2,3$. 
\cref{fig:TreePLS} (top row) displays the estimated covariate effect $\hat{C}$; the residual matrix $\hat{\Theta}$ and the corresponding edge probability matrix $\hat{P}$ obtained via the proposed iterative PLS algorithm. 
Visual comparisons of the estimated contribution of the covariates and the residual term to the overall network structure ($\hat{P}$), clearly suggest that the observed covariates contribute to the probabilities of tree interactions at a more local node-level rather than at a cluster level, with the underlying clusters being detected by the latent part of the model. Further, the plots allow us to identify regions (or tree-pairs) where covariates contribute significantly. For example, in this case simply as those regions where the positive values in the residual kernel $\hat{\Theta}_{ij}$ are of a lower intensity in comparison to the corresponding edge probability $\hat{P}_{ij}$. This is clearly observed, e.g. along the diagonal around $(30,30)$ where $\hat{P}\approx 1$, the residual kernel is $\approx 0.7$ and the covariate effect is $\approx 0.3$. 

\begin{figure}
\centering
\includegraphics[scale=0.35]{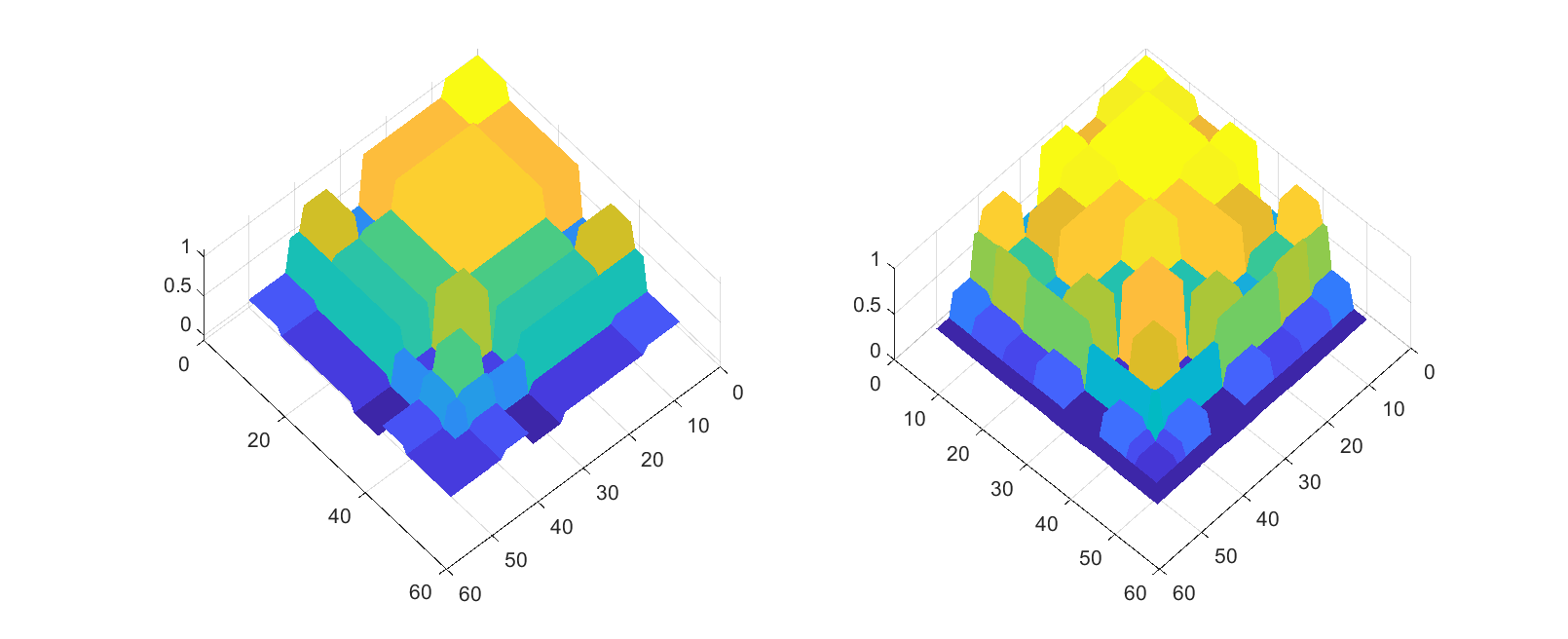}
\caption{\textit{Left}: Residual (latent) structure estimated using covariates via the proposed PLS; \textit{Right:} latent structure estimated without using covariates via ASE under the GRDPG model, for the tree network.}\label{fig:TreePLSSurf}
\end{figure}

\begin{figure}
\centering
\includegraphics[scale=0.58]{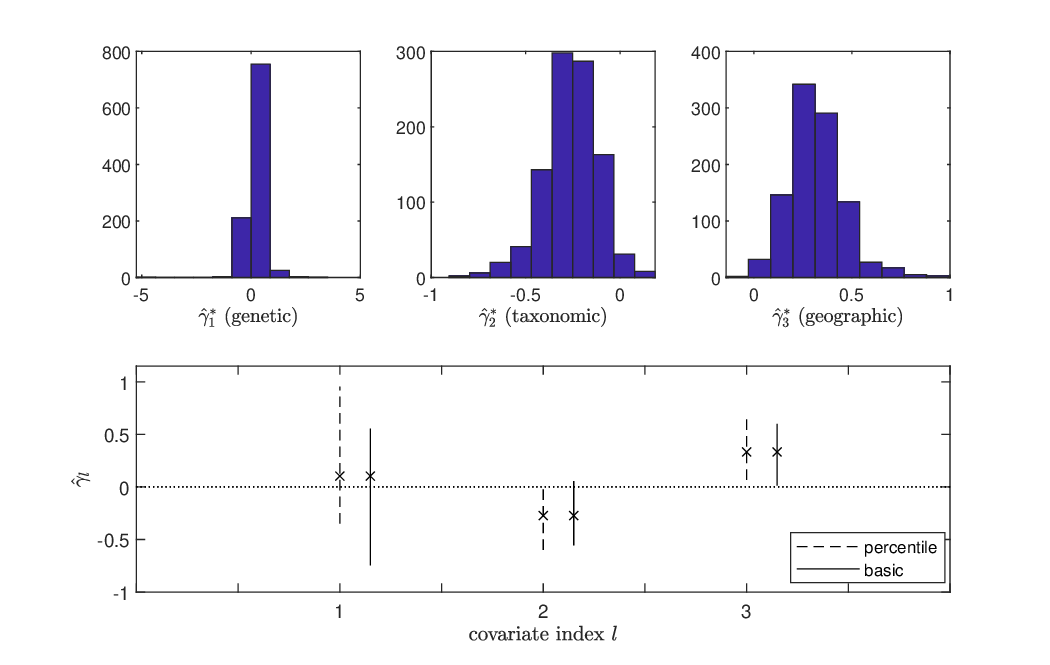}
\caption{Top row: Histograms of bootstrapped covariate coefficients $\hat{\gamma}^{\ast}_{l}, l=1,2,3$.
Second row: Estimated (edge) covariate coefficients (`x') $\hat{\bgamma}=[0.1032, -0.2721, 0.3332]^{T}$ with vertical bars displaying the corresponding $95\%$ percentile and basic bootstrap confidence intervals for $\gamma_{l}$.}
\label{fig:TreePLS_CI}
\end{figure}

Comparing the residual or latent structure estimated using covariates via the proposed method with the structure estimated without using covariates from GRDPG, \cite{Rubin2022} as displayed in \cref{fig:TreePLSSurf}, it is evident that many of the peaks and troughs in the GRDPG estimate disappear when latent structure is estimated with the inclusion of observed covariates through our approach. To quantify the similarity (or differences) in these two latent structure estimates, 
we computed the Normalized Mutual Information (NMI) and the Adjusted Rand Index (ARI) between residual clusters obtained from our method (with covariates) and the GRDPG clusters (without covariates). The NMI and ARI between the two sets of clusters are
$0.63$ and $0.44$, respectively, suggesting some similarity but also significant differences, and hence relevance of the observed covariates (e.g. \cite{Roy2019}).
To specifically examine the significance of covariates,
\cref{fig:TreePLS_CI} displays the estimated covariate coefficients $\hat{\gamma}_{l}$, l=1 (genetic), 2 (taxonomic), 3 (geographic), with the corresponding $95\%$ bootstrap percentile and basic confidence intervals for $\gamma_{l}$, \cite{Davison1997}. 
Both choices of bootstrap confidence construction, suggest that under our model, the geographic distances ($l=3$) are the most significant in explaining the observed shared fungal parasite based interactions between trees. 
Our findings are broadly aligned with analysis of the same network in \cite{Latouche15}.

\subsection{CKM physician friendship network}\label{sec:CKM}
We consider the physician friendship network created by \cite{Burt1987} from the data collected by \cite{CKM}
where each physician was asked to name three friends; and attribute information based on responses on $13$ aspects concerning the impact of network ties on the physicians' adoption of a new drug were recorded. 
No response was provided by physicians in some cases who were removed from our analysis leading to a network on $n=100$ physicians (nodes). Of the $13$ node covariates, we included (1) city of practice (4 values); (2) discussion with other doctors (3 values); (3) speciality in a field of medicine (4 values) and (4) proximity with other physicians (4 values) as categorical and the rest as quantitative covariates. As above, for any quantitative node covariate $x_{i}, x_{j}$ we constructed the corresponding edge covariate as the absolute difference $x_{ij}=|x_{i}-x_{j}|$; and following \cite{Latouche15} for any categorical node covariate $x^{\prime}_{i}, x^{\prime}_{j}$, we defined $x^{\prime}_{ij}=1$ if $x^{\prime}_{i}= x^{\prime}_{j}$ and set $x^{\prime}_{ij}=0$, otherwise. 

\begin{figure}
\centering
\includegraphics[scale=0.48]{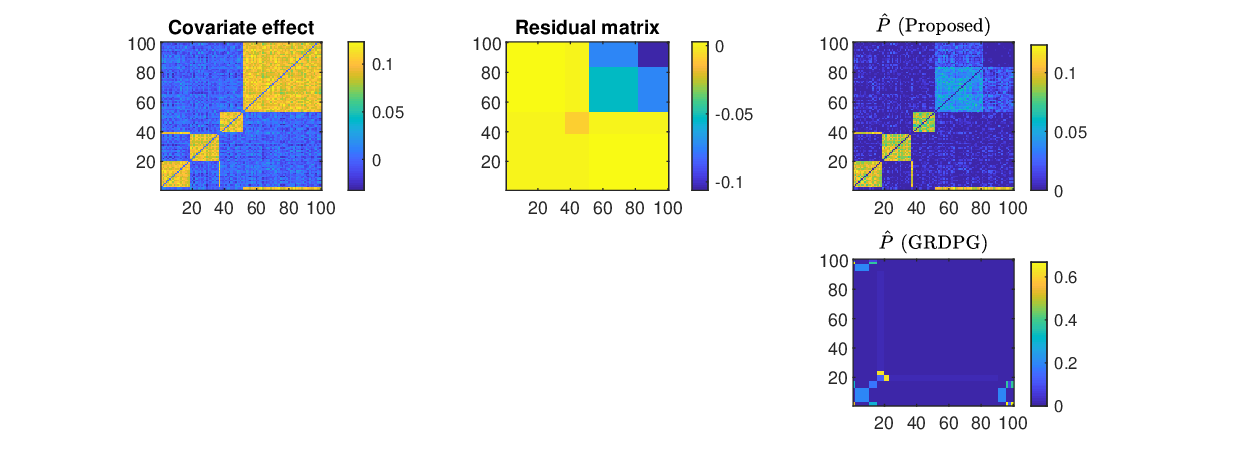}
\caption{\textit{First row}: Estimated covariate effect, residual structure, and the corresponding edge probability matrix $\hat{P}=[\hat{P}_{ij}]_{n \times n}$ via the proposed PLS; \textit{Second row}: $\hat{P}$ from ASE(A) under GRDPG, for the CKM network.}
\label{fig:CKMests}
\end{figure}

\begin{figure}
\centering
\includegraphics[scale=0.6]{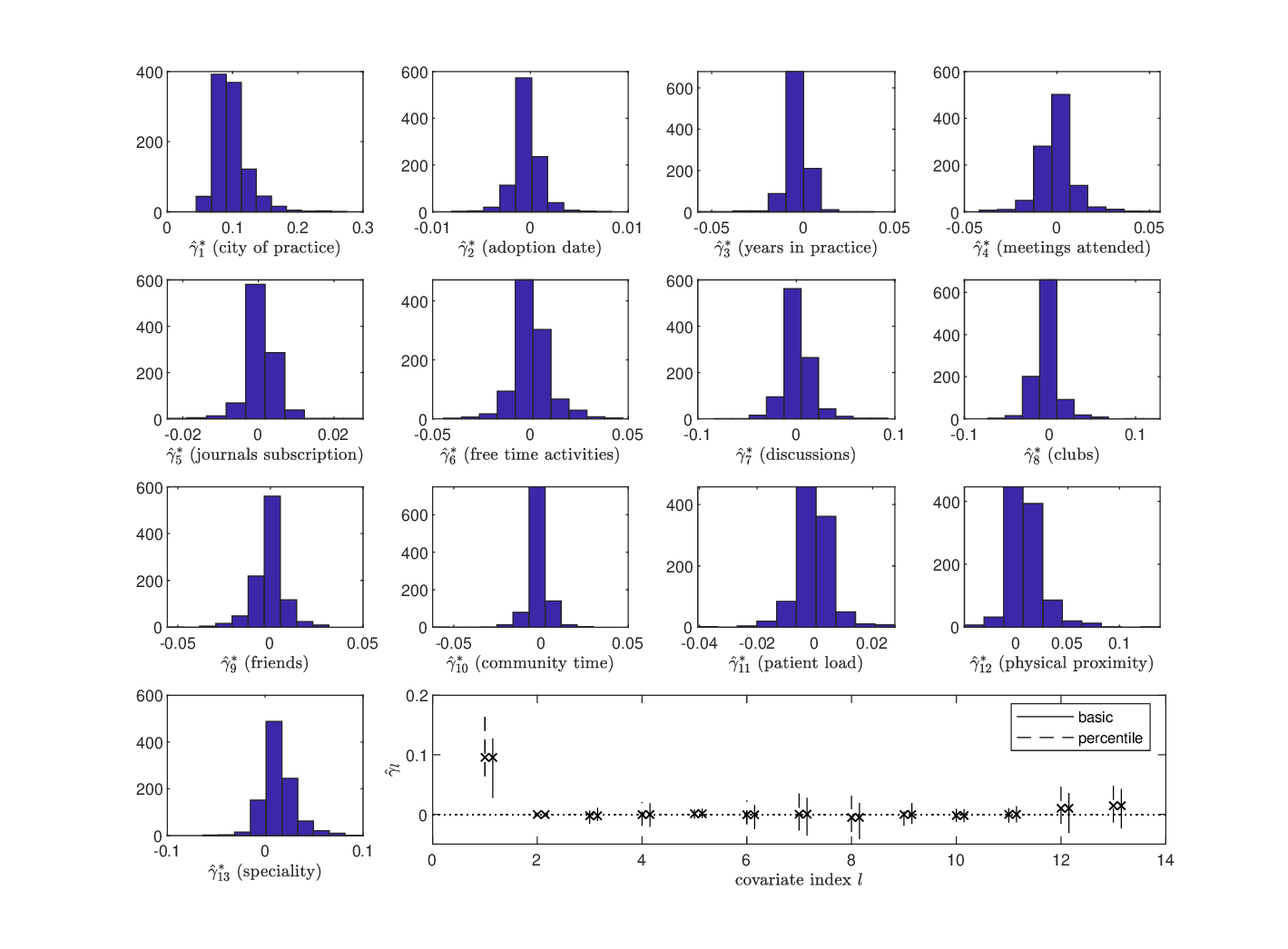}
\caption{Top row: Histograms of bootstrapped covariate coefficients $\hat{\gamma}^{\ast}_{l}, l=1,\ldots, 13$ Second row: Estimated (edge) covariate coefficients (`x') with vertical bars displaying the corresponding $95\%$ percentile and basic bootstrap confidence intervals for $\gamma_{l}$, for the CKM network.}
\label{fig:CKM_CIs}
\end{figure}

\begin{figure}
\centering
\includegraphics[scale=0.4]{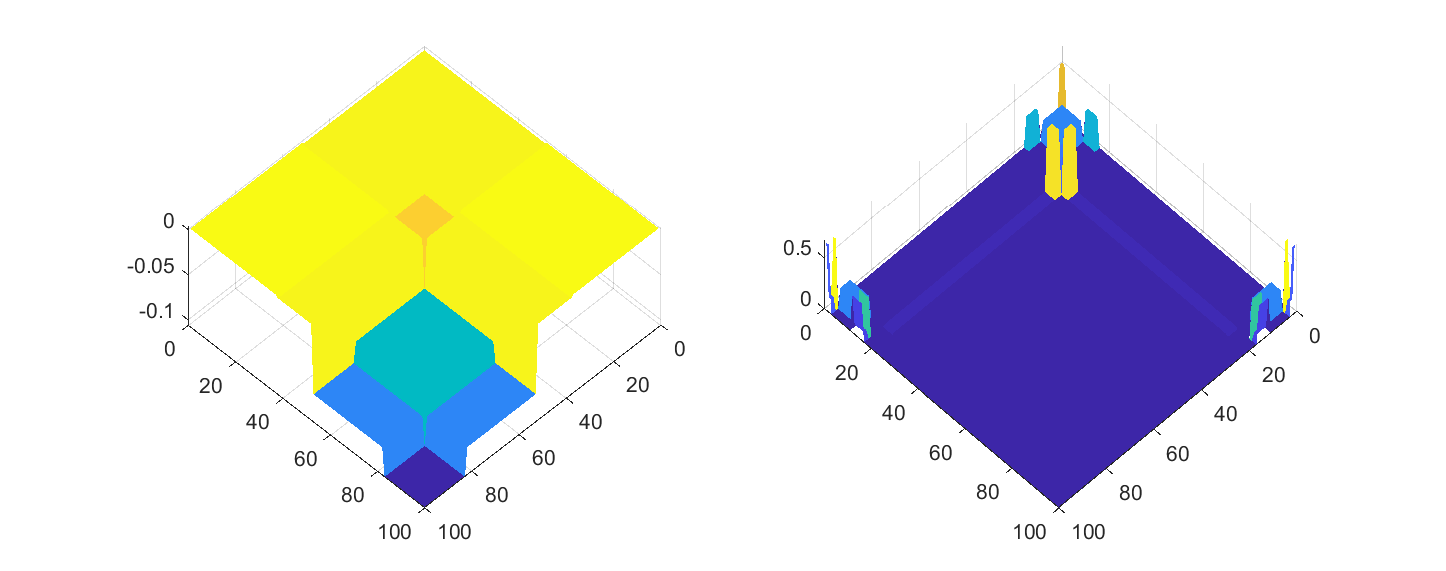}
\caption{\textit{Left}: Latent structure estimated using covariates via the proposed PLS; \textit{Right:} latent structure estimated without using covariates via ASE under the GRDPG model, for the CKM network.}
\label{fig:CKMsurfs}
\end{figure}

\cref{fig:CKMests} displays the covariate effect, residual matrix and the resulting edge probability matrix estimated via the proposed method and the edge probability matrix estimated from the adjacency matrix alone under the GRDPG model, \cite{Rubin2022}. Clearly $\hat{P}$ from the proposed method appears significantly different from that estimated under the GRDPG model. This is due to some or all of the observed covariates being extremely relevant in explaining the network structure as is evident from the similarity in structures of the $\hat{P}$ (proposed) and the covariate effect matrix displayed in \cref{fig:CKMests}. 
This is also reflected in the relatively low NMI and ARI scores of $0.32$ and $0.11$, respectively, observed between the residual clusters from our PLS approach (with covariates) and GRDPG clusters (without covariates). Visual comparisons with the GRDPG estimate suggest that the adjacency matrix alone is unable to capture the underlying structure in the network to the same extent in this case. 
It is also clearly seen from \cref{fig:CKMsurfs} that the residual latent structure as estimated from our method using covariates is mostly 0, suggesting that the covariates alone explain interactions for a large subset of node pairs in the CKM network.
To investigate the significance of covariates under our model, we obtained bootstrap samples $\hat{\gamma}^{\ast}_{l}, l\in[p]$ using the bayesian bootstrap as above.
\cref{fig:CKM_CIs} displays histograms of the bootstrapped covariate coefficients $\hat{\gamma}^{\ast}_{l}, l\in[p]$ with the final subplot displaying all estimates $\hat{\gamma}_{l}$, with the corresponding $95\%$ bootstrap percentile and basic confidence intervals for $\gamma_{l}$. Clearly, both choices of bootstrap confidence construction suggest that, under our model, only the city of practice covariate is most significant in explaining the network structure, consistent with existing studies in this area, \cite{MasciaEtAl2011,MasciaEtAl2015}.


\subsection{Military alliance networks} 
Building off of past network studies of military alliances and their determinants \cite{Warren2010,CranmerEtAl2012,MaozJoyce2016}, we applied our methodology to countries' military alliance networks for the years 1995 ($n=187$) and 2010 ($n=195$) in relation to the following $p=6$ covariates: (1) log material conflict (the log of the summed number of annual material conflict events between country pairs), (2) log of trade, (3) contiguous (a binary indicator of whether the pair of countries shared a land border), (4) civil conflict (a binary indicator of whether at least one of the two countries was experiencing a civil conflict), (5) joint democracy (a binary indicator of whether the two countries were jointly democratic), and (6) log of military expenditure. The three quantitative node covariates 1, 2, and 6 were converted to edge covariates via the absolute difference as in the CKM example above. The alliance data itself is taken from \cite{LeedsEtAl2002} and \cite{Leeds2018}.

\begin{figure}
\centering
\includegraphics[scale=0.5]{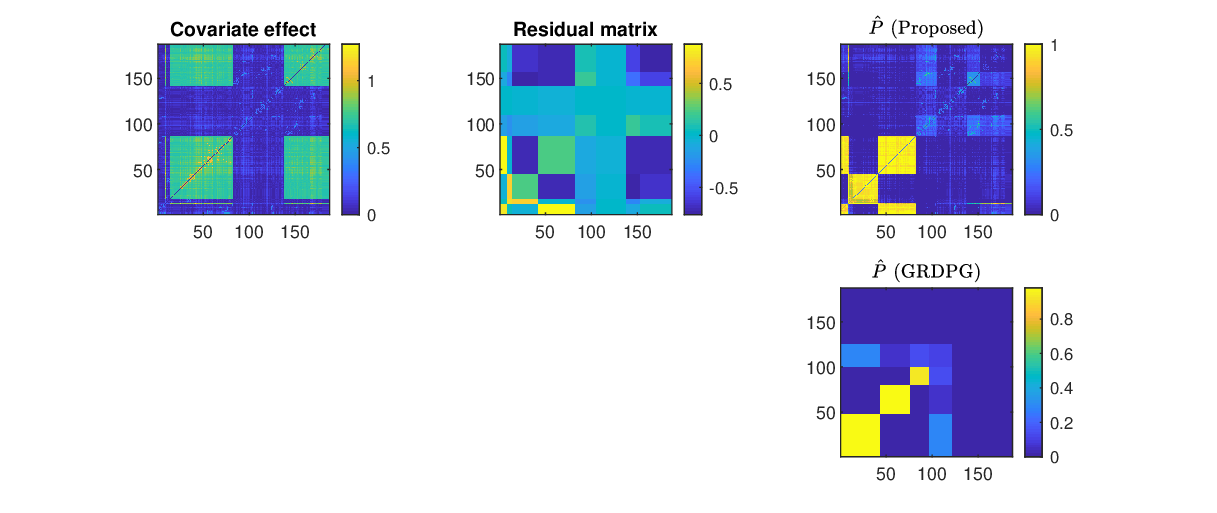}
\caption{\textit{First row}: Estimated covariate effect, residual structure, and the corresponding edge probability matrix $\hat{P}=[\hat{P}_{ij}]_{n \times n}$ via the proposed PLS; \textit{Second row}: $\hat{P}$ from ASE(A) under GRDPG, for the alliance network and covariate data in year 1995.}
\label{fig:Alliancet23ests}
\end{figure}

\begin{figure}
\centering
\includegraphics[scale=0.47]{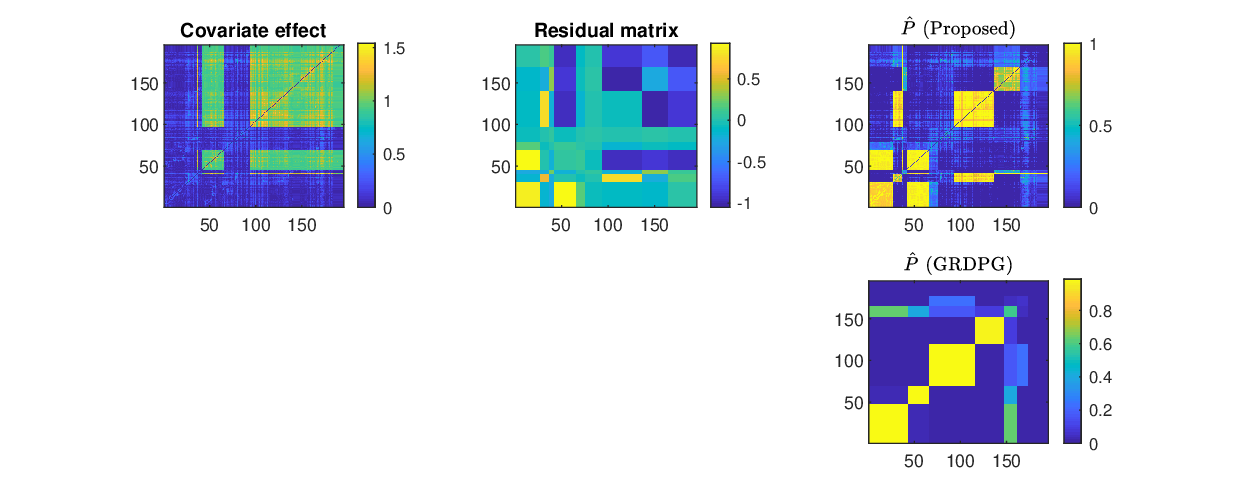}
\caption{\textit{First row}: Estimated covariate effect, residual structure, and the corresponding edge probability matrix $\hat{P}=[\hat{P}_{ij}]_{n \times n}$ via the proposed PLS; \textit{Second row}: $\hat{P}$ from ASE(A) under GRDPG, for the alliance network and covariates in year 2010.}
\label{fig:Alliancet38ests}
\end{figure}

The estimated structures displayed in \cref{fig:Alliancet23ests} and \cref{fig:Alliancet38ests} for years 1995 and 2010 illustrate how covariates contribute to a greater extent in explaining structure for some pairs of nodes than others and result in an edge probability estimator different to that estimated based on the adjacency matrix alone. Specifically, from \cref{fig:Alliancet23ests} (year 1995) we see that the first small high intensity block (primarily involving Russia and several Soviet states) near $(0,0)$ in $\hat{P}$ (proposed) is a contribution from the residual term whereas the remaining two high intensity clusters (along the diagonal of $\hat{P}$) are mostly contributions from the covariate term. These two clusters correspond to respective groupings of primarily Latin American/Caribbean, and European, country pairs for which the covariates contribute significantly to alliance formation. Likewise, from \cref{fig:Alliancet38ests} (year 2010), we see that among the five high intensity clusters along the diagonal with $\hat{P}$ (proposed) close to $1$, clusters 3, 4, and 5 result from the covariate term (with a close to zero residual), implying that alliances between the corresponding country pairs are almost fully explained by the linear covariate term. Interestingly, cluster 2---which aligns closely with the Soviet states identified in cluster 1 in 1995---now exhibits covariate contributions. This suggests that the (recently dissolved) Soviet Union's latent alliance pull in 1995 had subsided in favor of more standard edge and node alliance predictors by 2010. The broader split between the covariate and latent model components is again reflected in the NMI and ARI scores of $0.62; 0.72$ and $0.59; 0.65$, respectively, for the years 1995; 2010 between the residual clusters (with covariates) and GRDPG clusters (without covariates). This implies some overlap and some differences (due to significant contribution from covariates) between the latent structures as identified via the two approaches.  

\begin{figure}
\centering
\includegraphics[scale=0.52]{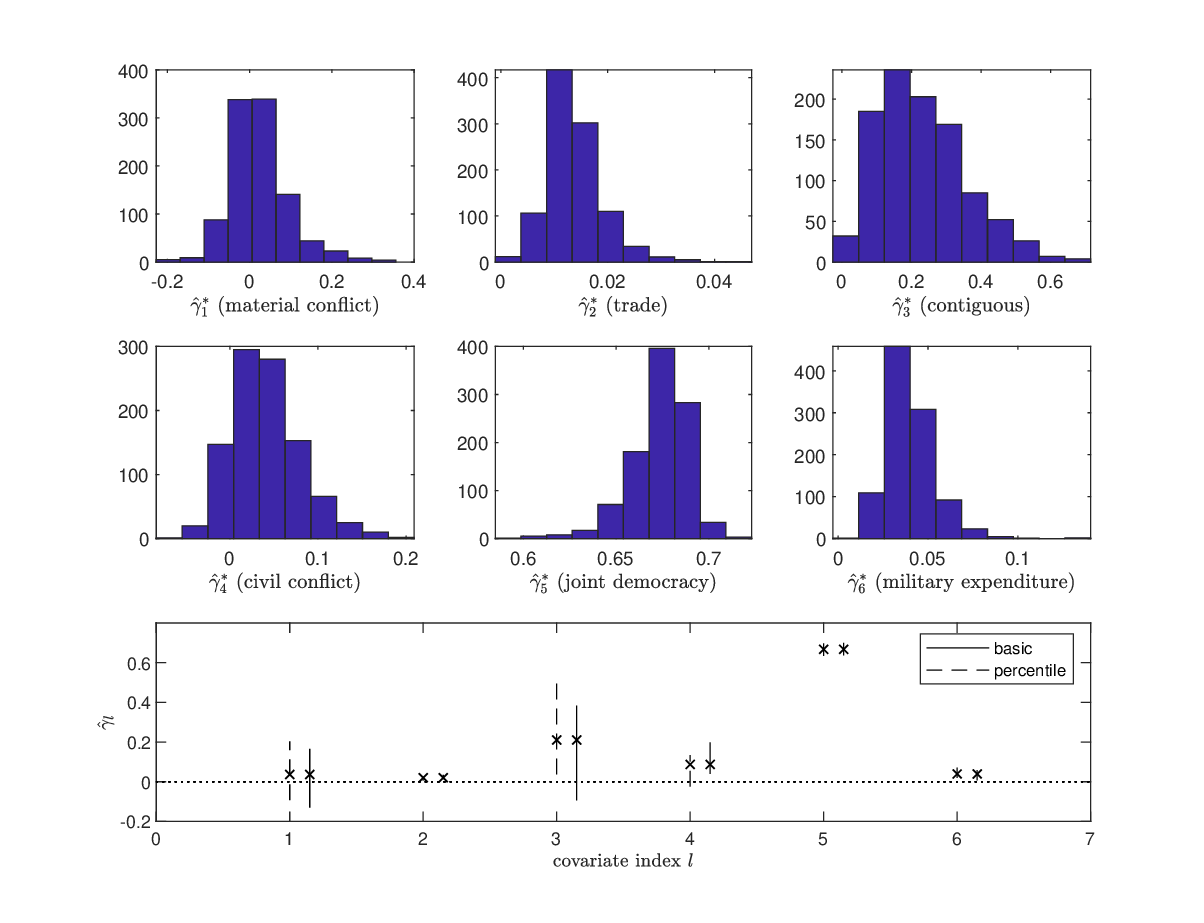}
\caption{Top row: Histograms of bootstrapped covariate coefficients $\hat{\gamma}^{\ast}_{l}, l=1,\ldots, 13$ Second row: Estimated (edge) covariate coefficients (`x') $\hat{\bgamma}=[0.037, 0.021, 0.21, 0.087, 0.667, 0.039]^{T}$ with vertical bars displaying the corresponding $95\%$ percentile and basic bootstrap confidence intervals for $\gamma_{l}$, for the alliance data in year 1995. Trade ($l=2$), joint democracy ($l=5$), and military expenditure ($l=6$) are significant.}
\label{fig:Alliancet23_CIs}
\end{figure}

\begin{figure}
\centering
\includegraphics[scale=0.52]{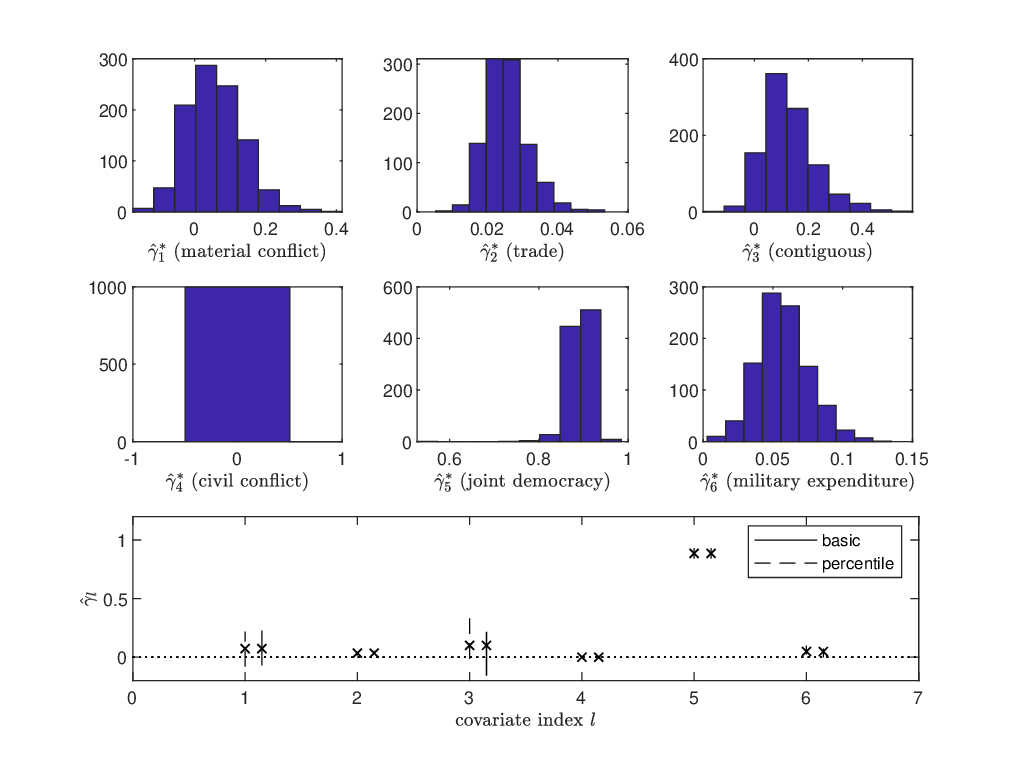}
\caption{Histograms of bootstrapped covariate coefficients $\hat{\gamma}^{\ast}_{l}, l=1,\ldots, 13$. Final row: Estimated (edge) covariate coefficients (`x') $\hat{\bgamma}=[0.073, 0.035,0.1,0,0.886,0.048]^{T}$ with vertical bars displaying the corresponding $95\%$ percentile and basic bootstrap confidence intervals for $\gamma_{l}$ for the 2010 alliance network.} 
\label{fig:Alliancet38_CIs}
\end{figure}

Bootstrap inference on our model's linear coefficients---as displayed in \cref{fig:Alliancet23_CIs} (year 1995)---suggest that the volume of trade between countries ($l=2$), joint democracy ($l=5$), and military expenditure ($l=6$) are significant in explaining alliances. Additionally, contiguous ($l=3$) and civil conflict ($l=4$) were found to be significant or not based on whether percentile or basic bootstrap construction were used, respectively. Bootstrap inference on the linear coefficients in year 2010 lead to similar conclusions with the same covariates ($l=2, 5, 6$) appearing significant based on the histograms as displayed in \cref{fig:Alliancet38_CIs}. Specifically, with the percentile confidence interval construction, trade ($l=2$), joint democracy ($l=5$), and military expenditure ($l=6$) are significant, whereas, with the basic construction, only, trade ($l=2$) and joint democracy ($l=5$) are significant. Note here that civil conflict coefficient is estimated to be exactly $0$ as the edge covariate data comprised of only $0$s in this year. 
The most consistent findings across 1995 and 2010---those of the alliance-inducing effects of joint democracy and trade---together help to reconcile what have been at times contradictory and counterintuitive network-oriented results in this vein \cite{Warren2010,CranmerEtAl2012} while reinforcing several past theoretical contentions \cite{Leeds1999,LaiReiter2000,Fordham2010}. Furthermore, in each year considered, significant positive values are observed in the residual matrix suggesting that (i) our covariates alone are not fully explanatory of alliances and (ii) latent  structure is key to modeling alliances. The latter insight reinforces similar contentions throughout past network-based studies of military alliances \cite{ParkSohn2020,DeNicolaEtAl2022}.

\section{Conclusion}
In this article, we address the problem of estimating network structure when node or edge covariates of mixed types--discrete, continuous, or both--are observed. Specifically, taking inspiration from classical semiparametric statistics, we consider the model which employs a linear covariate term and captures the remaining unobserved heterogeneity in network structure, through an indefinite inner product kernel of low-dimensional, latent node specific vectors. Noting why the profile least squares approach commonly used for classical  semiparametric models of similar type does not automatically extend to such network data, we proposed an iterative profile least squares algorithm for estimation of our network-covariate model. Additionally, we show how inference on the network model parameters can be drawn using the existing generalized bootstrap for estimating equations. Simulation study conducted on networks with different types of residual structures and covariate types confirmed the satisfactory performance of both our estimation algorithm and the bootstrap procedure. Application of our methodology to four real-world networks observed with covariates of different types and of varying sizes led to valuable insights on how network structure arises from the observed covariates (captured by contribution of the linear covariate term) and unobserved factors (represented by the residual term). The bootstrap-based inference procedure further enabled hypothesis testing on the significance of covariates in explaining network structure via the linear term. Thus, for networks observed with covariates, our method allows estimation of structure that is truly latent (via the residual term) and provides a powerful tool for understanding the 
 drivers of network structure.

\bibliographystyle{apalike}
\bibliography{graph_covar2_alliance}
\end{document}